\documentclass[iop]{emulateapj}

\usepackage{rotating}
\usepackage{amssymb}
\usepackage{amsmath}
\usepackage{color}
	
\shorttitle{Baryons Around Galaxies}
\shortauthors{Bregman et al.}

\begin{document}

\title{The Extended Distribution of Baryons Around Galaxies}

\author{Joel N. Bregman\altaffilmark{1}, Michael E. Anderson\altaffilmark{2}, Matthew J. Miller\altaffilmark{1}, 
Edmund Hodges-Kluck\altaffilmark{1}, Xinyu Dai\altaffilmark{3}, Jiang-Tao Li\altaffilmark{1}, 
Yunyang Li\altaffilmark{1,4}, and Zhijie Qu\altaffilmark{1}}

\email{jbregman@umich.edu}

\altaffiltext{1}{Department of Astronomy, University of Michigan, Ann Arbor, MI  48109}
\altaffiltext{2}{Max-Planck-Institut f{\"u}r Astrophysik, Karl-Schwarzschild-Str. 1, D-85748 Garching}
\altaffiltext{3}{Homer L. Dodge Department of Physics and Astronomy, University of Oklahoma, Norman, OK 73019}
\altaffiltext{4}{Department of Astronomy, Peking University, Beijing 100871, China}

\begin{abstract}

We summarize and reanalyze observations bearing upon missing galactic
baryons, where we propose a consistent picture for halo gas in L $\gtrsim$ L* galaxies. 
The hot X-ray emitting halos are detected to 50--70 kpc, where typically, 
M$_{hot}(<50 \textrm{ kpc}) \sim 5 \times 10^9$ M$_{\odot}$, and with density $n \propto r^{-3/2}$. 
When extrapolated to R$_{200}$, the gas mass is comparable to the 
stellar mass, but about half of the baryons are still missing from the hot phase. 
If extrapolated to 1.9--3R$_{200}$, the baryon to dark matter ratio approaches 
the cosmic value.  Significantly flatter density profiles are unlikely for 
R $< 50$ kpc and they are disfavored but not ruled out for R $> 50$ kpc. 
For the Milky Way, the hot halo metallicity lies in the range 0.3--1 solar for R $< 50$ kpc.
Planck measurements of the thermal Sunyaev-Zeldovich effect toward stacked luminous
galaxies (primarily early-type) indicate that most of their baryons are hot, near the virial temperature, 
and extend beyond R$_{200}$.  This stacked SZ signal is nearly an order of magnitude larger 
than that inferred from the X-ray observations of individual (mostly spiral) galaxies 
with M$_* > 10^{11.3}$ M$_{\odot}$. This difference suggests that the hot halo properties
are distinct for early and late type galaxies, possibly due to different evolutionary histories.
For the cooler gas detected in UV absorption line studies, we argue that there are two absorption 
populations: extended halos; and disks extending to $\sim 50$ kpc, containing most of this gas, 
and with masses a few times lower than the stellar masses. 
Such extended disks are also seen in 21 cm HI observations and in simulations.

\end{abstract}



\section{Introduction}

During the past two decades, we have come to understand that galaxies are
baryon-poor (e.g., \citealt{moster10,mcgaugh10,Dai2012}).  
That is, the dynamics of a galaxy (rotation curves or velocity
dispersion) is interpreted within the framework of the NFW \citep{navarro97} distribution to
define its mass.  This mass of baryons originally associated with this dark
matter halo is given by the dark matter to baryon ratio that is known to
high accuracy from the CMB observation, about 5.3:1 \citep{planckXVI}. 
The easily observed baryonic mass, the stars and cool gas, is significantly
lower than the pre-collapse mass, by factor of 2--100 \citep{mcgaugh10}. 
Evidently, the act of galaxy formation led to a fraction of the baryons
falling deep into the potential well, becoming the familiar galaxies of
today.
%% general introduction to the subject

This situation has motivated a great deal of theoretical and observational
work, with many of the observational efforts directed toward discovering
the location and properties of the baryons that are ``missing" from galaxies
today (e.g., \citealt{gou2010,dave11,pion11,scan12,voge14,scha15}).  
One prediction for the missing baryons is that they reside in a hot
state around galaxies (e.g., \citealt{white78}, \citealt{white91}, \citealt{fuku06}), with an extent as great or
greater than the dark matter halos.  This gas acquires a temperature
comparable to the dynamical temperature of the system by a combination
of shock heating associated with infall plus heating from supernovae and
AGN (e.g., \citealt{crain2015}).  The relative contributions of these heating agents is model
dependent, but each should leave different signatures, which involve the
gas masses at different temperatures, the radial distributions of the gas
components, as well as the metallicity distributions.
%% first discussion of the halos, hot and otherwise

Observational efforts to study hot halos necessarily involve X-ray data, as the 
halos should be near their virial temperatures, $\gtrsim 10^6$ K for an $\gtrsim$L* galaxy, 
where nearly all of the important lines occur at X-ray energies.  X-ray absorption 
line observations are confined to studies of the Milky Way halo, but emission 
line investigations address both the Milky Way and external galaxies \citep{osull2007, 
ander2010, ander2011, ander2013, bogdan13, miller2013, bogdan15, walker15, miller2015, ander2016}. 
Observations of gas well below the virial temperature are commonly seen in galaxy 
halos (e.g., \citealt{putman12}), at $\sim 10^4$ K for most of the UV absorption line gas, which is modeled 
as photoionized clouds (e.g., \citealt{werk2014}). Some of the clouds of higher 
ionization state ions, notably O VI, may also be produced in collisional ionization 
from gas near $10^{5.5}$ K (e.g., \citealt{stocke2014}).
These gaseous components  cannot be in hydrostatic equilibrium on the scale 
of $R_{200}$, so one expects the gas to fall to the disk on a relatively short 
timescale of about 1 Gyr.  That would suggest that this gas is not a major mass 
component of the halo, but some results indicate otherwise \citep{werk13a}.
%% more detailed discussion of observations

One scientific goal of these observational programs are to obtain the density
distribution of the hot gas, from which the gas mass can be determined.
Another important goal is to measure the temperature of the hot halo, which 
reflects the effects of infall and feedback (e.g., \citealt{fielding16, zhijie2018}).  
Feedback from stars is responsible for the metallicity in the halo, also of critical interest.
Finally, the dynamics of the hot gas informs us of the infall or outflow, the
turbulence, and the rotation of the hot halo, if present.

A variety of observational programs have made progress in these areas and our
goal is to synthesize these results into a coherent picture. 
In the first part of this paper, we consider the hot gas in the Milky Way and 
external galaxies ($\S$2), with an analysis of the fraction of baryons within $R_{200}$.  
We also reconsider the importance of the cooler
gas, seen in absorption in the UV ($\S$3).  Related observations bear
on these issues, such as the Sunyaev-Zeldovich measurements from Planck
($\S$4), which are sensitive to hot gas around the galaxies
\citep{planckXI}, and to metallicity issues.  
We conclude by summarizing the current state of halo gas distributions, 
identifying areas of consistency and stress between investigations, 
arguing for a disk-halo model for extended gas distributions, and suggesting
future observations ($\S$5).

\section{Hot Gas Density Distributions around galaxies: Implications for Missing
Baryons}

For the analysis of hot gas, one usually assumes that the gas is near hydrostatic equilibrium at or above the virial temperature.  When the temperature varies significantly less than the density, the gas density has a power-law dependence on radius beyond the core radius.  
This class of density profiles, the $\beta$-model \citep{cav1976}, has 
the functional form $n = n_o/(1+(r/r_c)^2)^{3{\beta}/2}$. 

\begin{figure}       %%%%%%%%%%%%% Figure 1
\centering
%\epsscale{0.8}
\plotone {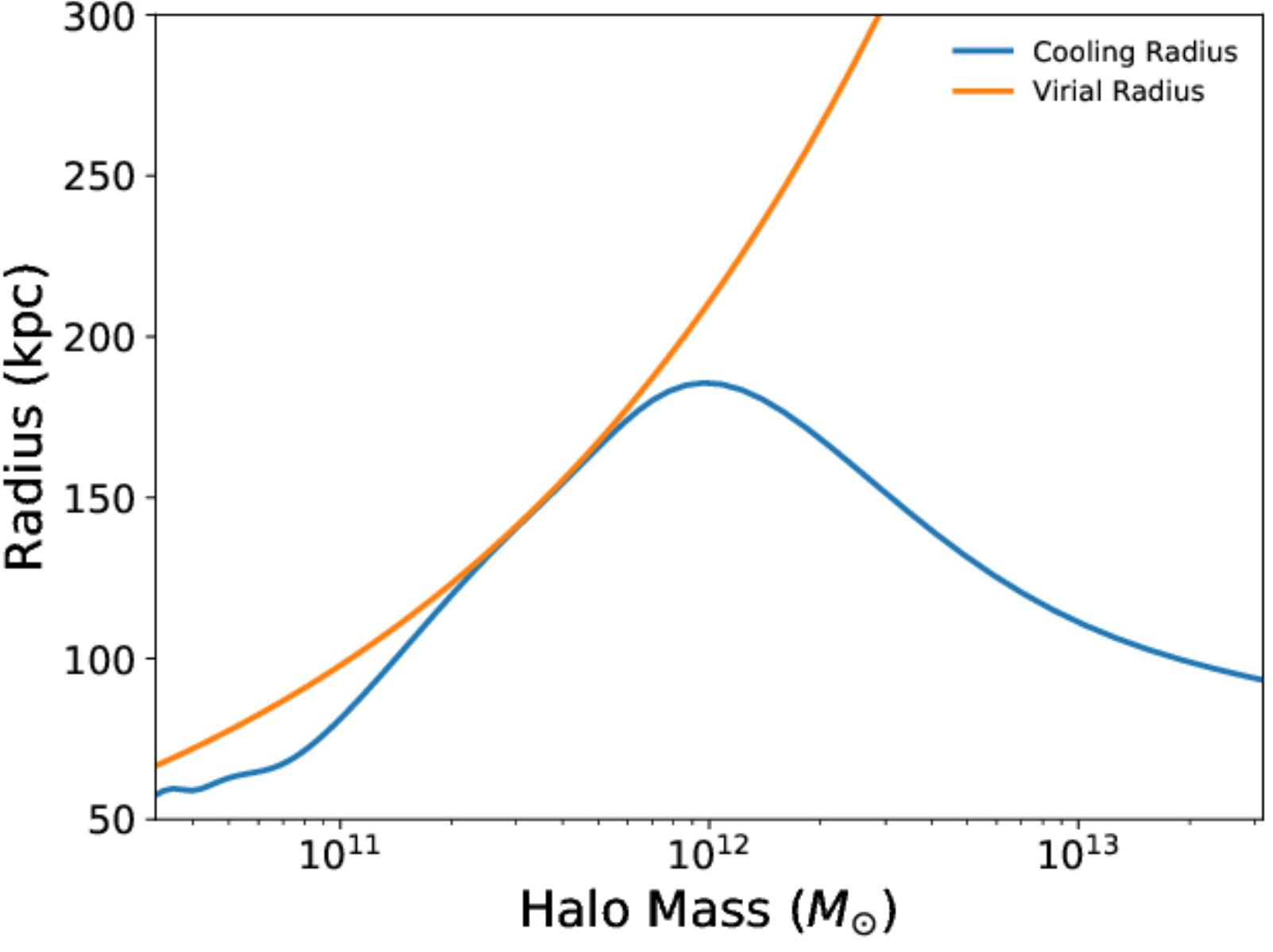}
\caption{
The virial and gaseous cooling radius as a function of gravitating halo mass for model {\tt TCPIE} from \citet{zhijie2018}.  The range of the cooling radius lies in the 60-190 kpc range over a three order of magnitude range in gravitating halo mass. 
}
\label{fig:CoolingRadius}
\end{figure}

\begin{figure}       %%%%%%%%%%%%% Figure 2
\centering
%\epsscale{0.8}
\plotone {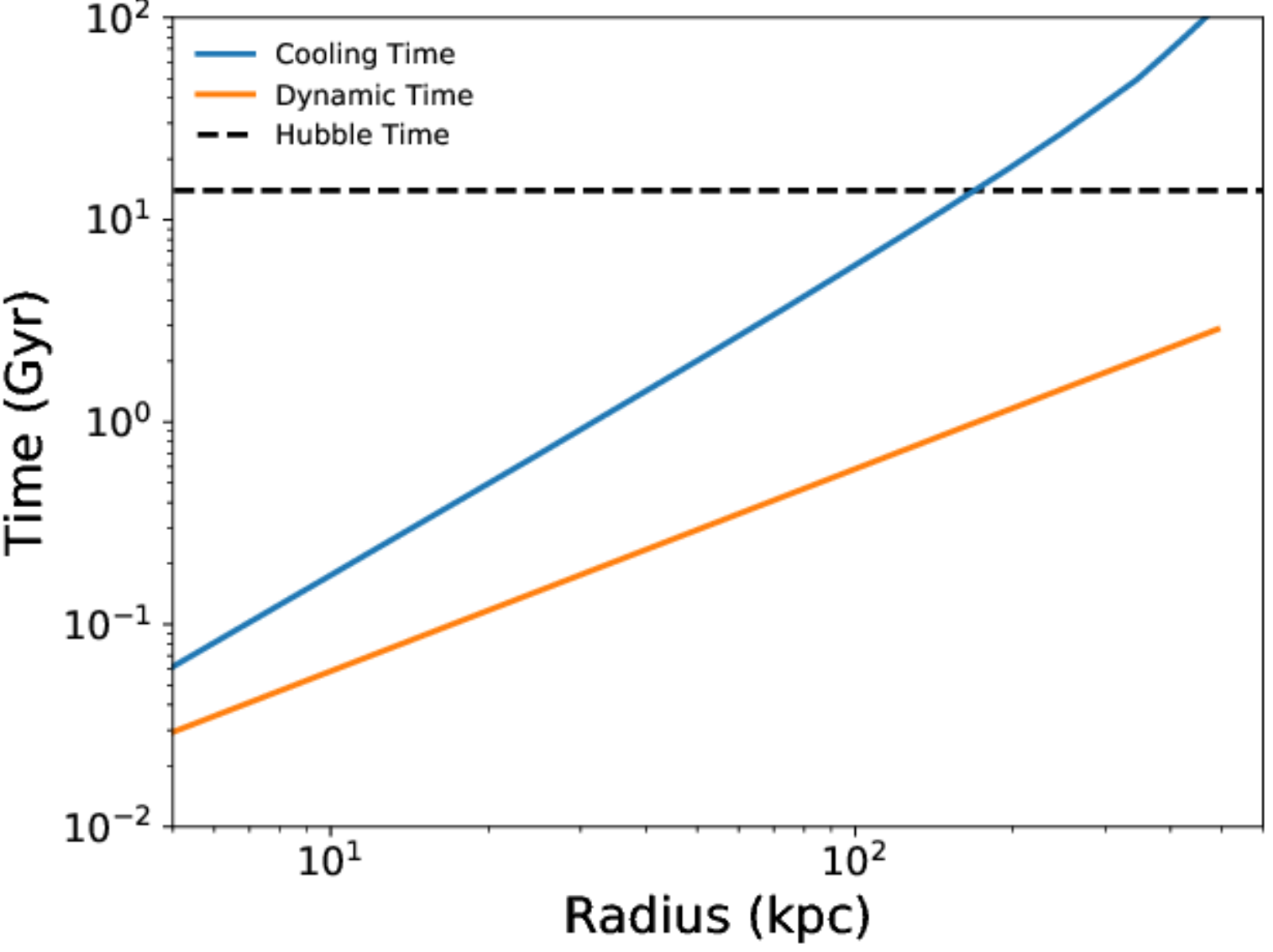}
\caption{
The cooling and dynamical (free-fall or sound-crossing) times as a function of radius for a Milky Way type galaxy where logM$_{halo}$ = 12.2, logM$_*$=10.8, R$_{vir}$ = 246 kpc, and T$_{gas}$ = $2 \times 10^6$ K ({\tt TCPIE}, after \citealt{zhijie2018}).  The cooling time, which reaches a value of 10$^9$ yr at 30 kpc, is always longer than the dynamical time.  The gaseous halo extends to 2 R$_{vir}$ in the above model.
}
\label{fig:CoolingTimes}
\end{figure}

To consider the validity (and other aspects) of such a model, we developed a semi-analytic spherically symmetric model that has a hot gaseous halo near the virial temperature \citep{zhijie2018}.  The ionization state of the gas is modified by photionization from the metagalactic radiation field \citep{haardt2012}, which changes the cooling function.  A value of $\beta = 1/2$ is adopted, leading to  $n \propto r^{-3/2}$ beyond the core radius. 
A cooling radius is defined in the usual way where the cooling time equals the Hubble time, and within this cooling radius, the gas is assumed to cool at a rate equal to the star formation rate, which is given as a function of galaxy stellar mass by the star formation main sequence of star-forming galaxies \citep{moreselli16}.  For a metallicity of 0.5 solar, the cooling radius occurs in the 60-190 kpc range (Figure ~\ref{fig:CoolingRadius}) over a wide range of galaxy masses for cooling flow models with feedback where collisional ionization equilibrium is modified by photoionization (after model {\tt TCPIE} of \citealt{zhijie2018}).  Also, the mass of the hot halo out to $2R_{200}$ increases approximately with the gravitating halo mass and is often comparable to and sometimes greater than the stellar mass.  The radiative cooling time is longer than the free fall or sound-crossing time (similar values), supporting the use of  the hydrostatic assumption (Figure ~\ref{fig:CoolingTimes}).  Also, a typical accretion velocity is 20 km s$^{-1}$ \citep{miller2016a}, which is well below the sound speed of the gas, so ordered accretion does not have a significant effect of the density structure. We discuss suggested variations in the density law below.

There is a relationship between the temperature and $\beta$, and when the temperature 
varies less rapidly than the density and in the absence of turbulence and, $\beta =  T_{rot}/T$, 
where $T_{rot}$ is the thermal energy associated with the circular rotational velocity 
$v_{rot}(r) = (GM(r)/r)^{1/2}$, which is similar to the virial temperature but varies with radius.  
A model with $\beta = 1/2$ implies $T = 2 T_{rot}$, 
although turbulent energy will also contribute, so we might expect that $T \leq 2 T_{rot}$, 
which is consistent with observations, as discussed below.

\subsection{The Milky Way Halo Gas Density Distribution from X-Ray Absorption Line Data}

At temperatures of $1-2 \times 10^6$ K, the most important lines originate from the He-like 
and H-like oxygen ions (O VII and O VIII), with the O VII He$\alpha$ resonant line at 21.60 \AA\ 
being the strongest absorption line, followed by the O VIII Ly$\alpha$ resonant line at 
18.97 \AA , which has a fractional equivalent width that is about five times weaker
for the same ionic column densities.  
The O VII ion is present over a broad temperature range, $5.4 <$ logT $< 6.5$ 
(from the peak ion fraction to an order of magnitude below the peak)
while the O VIII ion is most common in the temperature range  $6.1 <$ logT $< 6.8$.
These and other lines have been detected in \textit{XMM-Newton} and \textit{Chandra} 
X-ray grating spectra against the continuum of bright background AGNs \citep{nica02, 
rasm2003, williams05, miller2013, fang15, neva2017}.  
Here we concentrate on the O VII He$\alpha$ line values, which have the most detections
and highest S/N of any X-ray absorption lines.

There are few sight lines that pass through the bulge region, so constrains on
the core radius have been poor when fitting a $\beta$ model.  
One can either fix the core radius at a value typical for 
early-type galaxies (1--3 kpc) or use the form of the $\beta$ model where $r >> r_c$,

$n(r)=\frac{n_{0}r_{c}^{3\beta}}{r^{3\beta}}$.

\citet{miller2013} used the latter method, although both give the same results, 
within the uncertainties.  For an optically thin plasma, N(O\ VII) = 
$3.48 \times 10^{14} EW$ cm$^{-2}$, where the equivalent width (EW) is in m\AA .  
The fitting leads to best-fit $\beta$ values of $0.56^{+0.10}_{-0.12}$ 
if the lines are optically thin and $0.71^{+0.13}_{-0.14}$ if the lines 
are mildly saturated (\cite{miller2013}, assuming a Doppler width of 
150 km s$^{-1}$ for all observations.
This resulted in saturation correction factors of $\approx 1-2$, at about a 
the $3\sigma$ level.

Using more recent data, \citet{fang15} assembled a larger sample of O VII absorption
line measurements and found a correlation of the EW versus angle from the Galactic 
Center for targets with $|b| < 45^{\circ}$ at the 95\% confidence level.
However, they state that they do not find a strong correlation of equivalent widths with
Galactic coordinates.  As this would seem to conflict with
the findings of \citet{miller2013}, we examined whether it yields a significantly
different result when fitting a $\beta$ model to their data set. 

The data set of \citet{fang15} consist of 33 O VII equivalent width measurements from 
43 sight lines. We exclude the ten sight lines where no significant absorption 
was detected (reported as 3$\sigma$ upper limits).  The authors do not discuss whether 
these are primarily due to low continuum S/N or weak absorption features, although 
there are several indications that the former causes these non-detections.  
Many of these sight lines are projected near other sight lines with significant O\ VII detections.  
This implies these sight lines should have detectable absorption if the absorption 
signature varies smoothly across the sky.  Moreover, 8/10 of the non-detections occur 
in sight lines with counts per resolution element below the median sample value.  
Thus, the non-detections are likely due to low S/N spectra and excluding them should 
not bias our model fitting results.  

Our hot gas density model and fitting procedure follow previous conventions discussed above.  
The hot gas electron density model is a modified $\beta$ model defined as a power law 
extending to the virial radius as given above, where $n_{0}r_{c}^{3\beta}$ is the 
normalization and $3\beta$ is the density slope.  

We used a Markov chain Monte Carlo (MCMC) algorithm to explore the model 
parameter space and find a best-fit model.  This code maximizes 
the likelihood between the model and data, where we define $ln(L) = -0.5\,\chi^2$.  
Thus, our best-fit model maximizes the likelihood and minimizes the $\chi^2$.  
We bin the output chains from the MCMC code and treat these as probability density 
functions (pdfs) for each model parameter.  The shapes and locations of the density 
function define the best-fit density model.

\begin{figure}       %%%%%%%%%%%%% Figure 3
\centering
\plotone {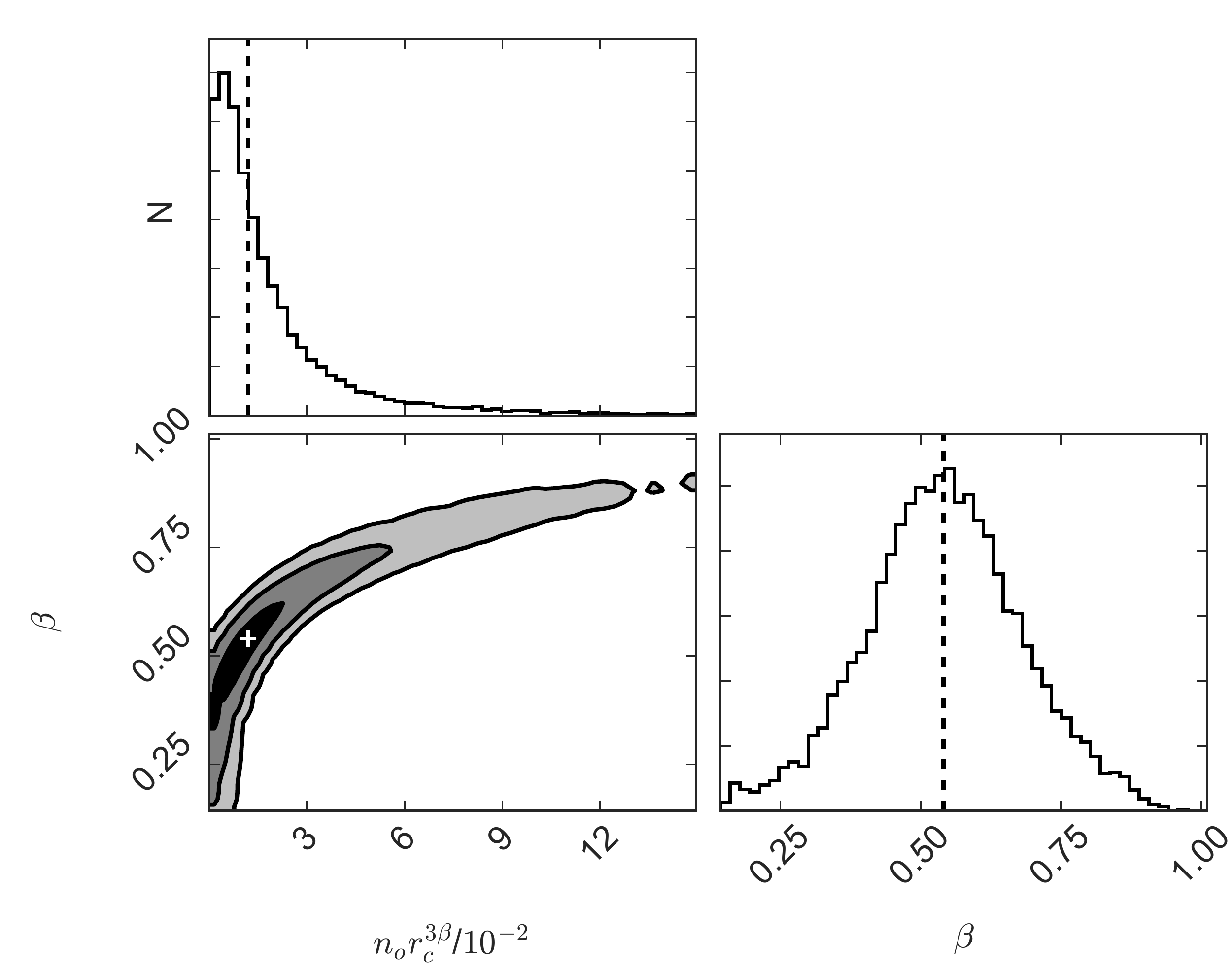}
\caption{
Probability density functions and contour plot for our hot halo model parameters based
on the data set of O VII absorption lines from \citet{fang15}.  The vertical dashed lines 
and white cross represent the median of each distribution, which we define as our best-fit model.  
The contour shades from black to gray represent the 1$\sigma$, 2$\sigma$, and 3$\sigma$ regions.  
The results are the same as those determined by \citet{miller2013} and \citet{hk2016}.
}
\label{fig:fang}
\end{figure}

Our results are seen as pdfs and a contour plot in Figure ~\ref{fig:fang}.  
We define the best-fit model as the median value of each parameter pdf and give $1\sigma$
uncertainties as the 68\% range away from the median value.  
The best-fit density model has parameters of 
$n_{0}r_{c}^{3\beta} = 1.20_{-0.82}^{+2.13} \times 10^{-2}$ cm$^{-3}$ kpc$^{3\beta}$ 
(at solar metallicity and $\beta = 0.54^{+0.14}_{-0.13}$).  
Similar to \citet{miller2013}, these results include 
an additional uncertainty of 7.5 m\AA\ added in quadrature to the observed equivalent 
widths to find an acceptable $\chi^2$ (reduced $\chi^2$ = 1.4 with 30 dof).  
These results are also consistent with the aforementioned study by \citet{miller2013}, who
found $n_{0}r_{c}^{3\beta} = 1.30^{+1.60}_{-1.00} \times 10^{-2}$ cm$^{-3}$ kpc$^{3\beta}$ 
(for solar metallicity), and $\beta = 0.56^{+0.10}_{-0.12}$. 

More recently, \citet{hk2016} compiled an updated set of O VII equivalent widths for a study
of Galactic rotation. They fit a disk plus halo gas model, where the disk component made a 
10\% contribution and led to a halo component with the parameters 
$n_{0}r_{c}^{3\beta} = 1.43 \pm 0.25 \times 10^{-2}$ cm$^{-3}$ kpc$^{3\beta}$ and
$\beta = 0.53 \pm 0.03$, which is a significant improvement on the accuracy of the 
density normalization.
To conclude, $\beta$-model fits to the samples of \citet{miller2013}, \citet{fang15}, and
\citet{hk2016} are indistinguishable from each other and support a 
radially decreasing density profile with $\beta \approx 0.5$ for the optically thin case.

\subsection{The Milky Way Halo Gas Density Distribution from X-Ray Emission Line Data}

The absorption line data set contains 2--3 dozen useful sightlines, but the 
emission lines data sets for Milky Way is about 1800 sightlines 
\citep{hs12,hs13}, which lead to stronger constraints on the density profiles. 
For our analysis, we chose a subset of sightlines
that avoid known bright objects (e.g., SNR, clusters of galaxies) and avoid observations
that might have problematic solar wind charge exchange contributions; this results in 
648 sightlines for which both O VII and O VIII emission is available \citep{miller2015}.
\citet{miller2015} considered the optically thin case and estimated a correction for optical 
depth effects.  

In a recent work, we include radiative transfer effects for the O VII He$\alpha$ 
triplet and the O VIII Ly$\alpha$ lines by using a Monte-Carlo radiative transfer 
model \citep{yunyang2017}.  Both a non-rotating halo and a rotating halo 
(v$_{\phi}$ = 183 $\pm$ 41 km s$^{-1}$; \citealt{hk2016}) were considered, along 
with models that included disk components.  
The best-fit model includes rotation and a disk component, although the disk component is a minor mass component, as found previously; the O VII and O VIII fits yield the same results. 
This analysis was able to constrain the core radius, so that $r_c = 2.53 \pm 0.18$ kpc, 
which is consistent with the separate analysis of the inner part of the Galaxy and the 
Fermi Bubbles by \citet{miller2016b}.  The slope of the density distribution is 
$0.51 \pm 0.02$ with a normalization of $n_{0}r_{c}^{3\beta}= 2.82 \pm 0.33 \times 10^{-2}$ cm$^{-3}$ kpc$^{3\beta}$ (for a metallicity of 0.3 solar). 
Turbulence or motion is implied by the non-thermal component of the Doppler $b$ parameter, where $b_{turb} = 110 \pm 45$ km s$^{-1}$.  
For the disk component, the best-fit vertical exponential scale height is $z_h = 1.34 \pm 0.47$ kpc and a radial scale length of 3 kpc was assumed. 
The exponential disk mass, $1.4 \times 10^8 M_{\odot}$ is small compared to the 
hot halo mass of $3.1 \times 10^{10} M_{\odot}$ within 250 kpc \citep{yunyang2017}.

The fits to the emission line data do not put useful constraints on the
metallicity, but constraints can be obtained when comparing the emission to
the absorption line data. \ That is because the emission depends on the
integrated emission measure, $Zn_{e}^{2}$, while the absorption depends on the
integrated column, $Zn_{e}$. In principle, this permits one to solve for the
metallicity $Z$, but in practice a joint fit is difficult because the
statistical power of the emission line fits dominates a joint fit. \ Instead
of a joint fit, we calculate model equivalent widths from the emission line
models for different values of the metallicity; opacity effects are included. 
The equivalent width sample taken from \citet{hk2016}, which has 37 sight lines, from which
we used only those sources where the S/N $> 10$ in the continuum near the O VII line.
This removes low S/N measurements with large errors, providing a sample of 26 lines of sight.

\begin{figure}       %%%%%%%%%%%%% Figure 4
\centering
%\epsscale{0.8}
\plotone {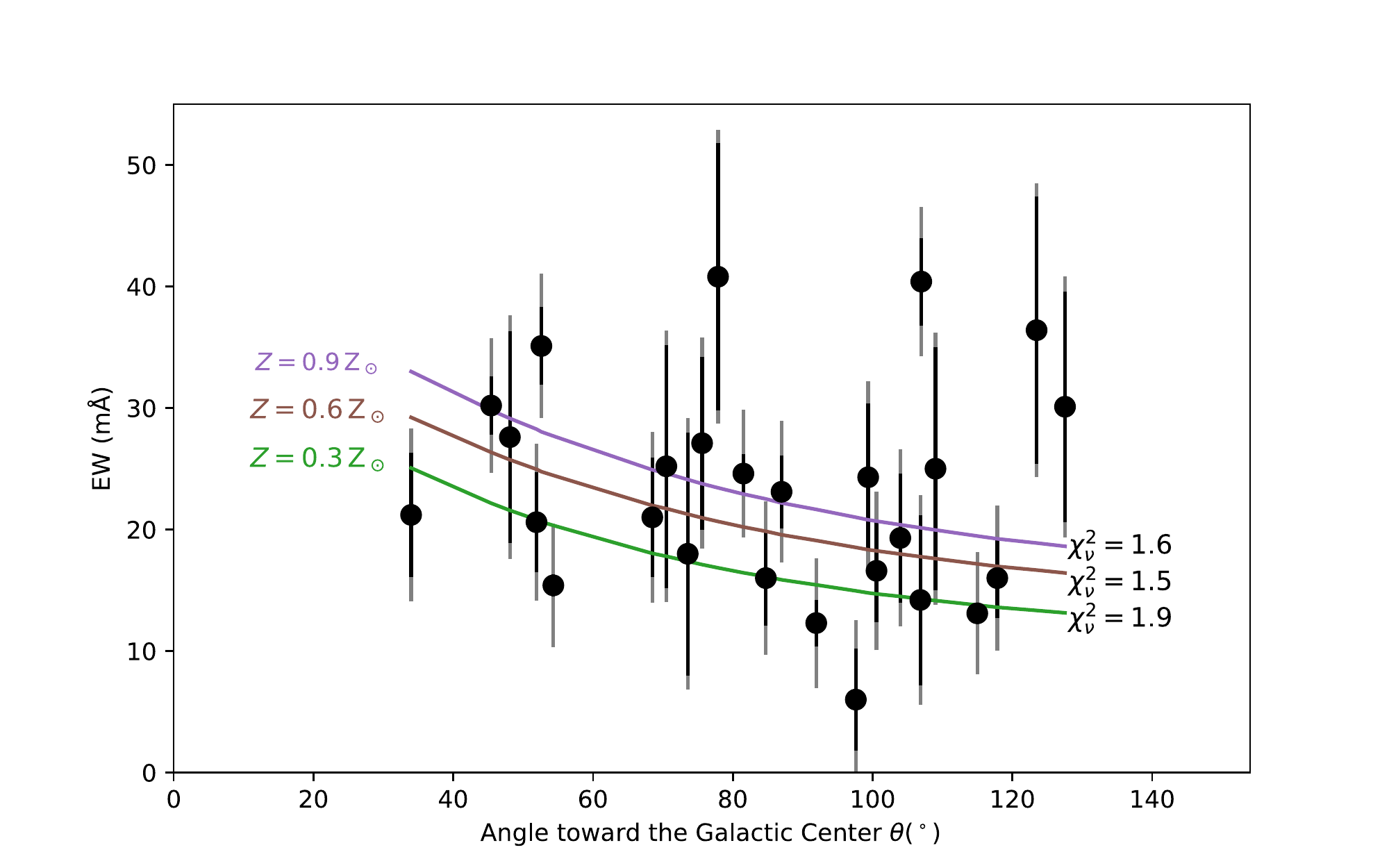}
\caption{
The distribution of O VII absorption line equivalent width observations \citep{hk2016},
with the addition of a 7.5 m\AA\ line of sight uncertainty, is compared to the 
values inferred from the fits to the O VII and O VIII emission line measurements \citep{yunyang2017}
for different values of the metallicity (solid lines). 
}
\label{fig:Z1}
\end{figure}

\begin{figure}       %%%%%%%%%%%%% Figure 5
\centering
%\epsscale{0.8}
\plotone {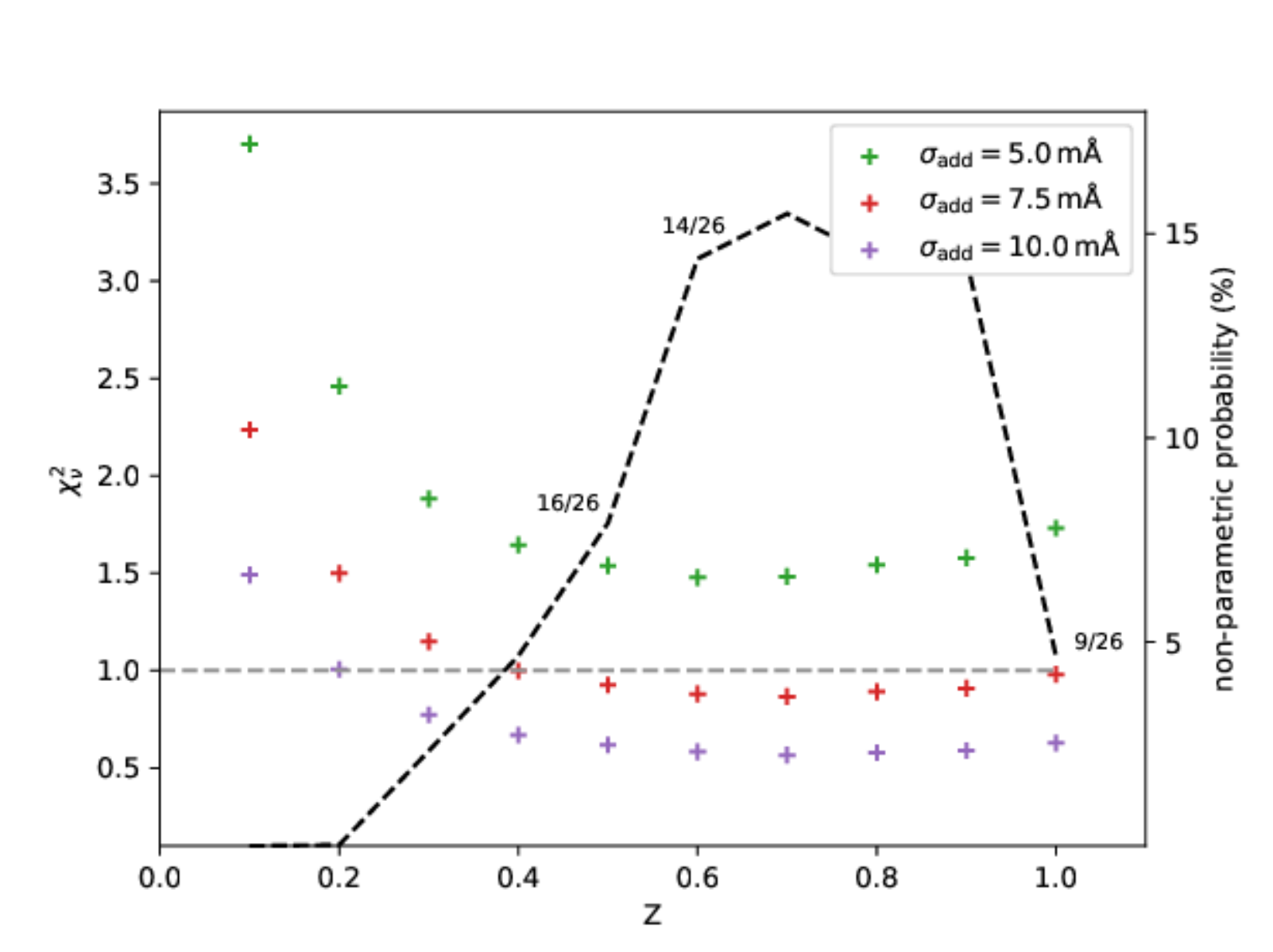}
\caption{
The quality of the agreement between the O VII absorption and emission line data, 
as a function of metallicity and for two different values of the additional line 
of sight variation in the equivalent widths ($\sigma_{add}$).  
The crosses represent $\chi_{\nu}^{2}$ values (left scale) in fits to the emission lines
while the black dashed line shows the non-parametric probability (right scale), based on
the number of points that lie above the model (numbers given in at three points).
For metallicities $Z < 0.3$, the $\chi_{\nu}^{2}$ value rises significantly and
nearly all of the equivalent width values lie above the model, indicating that $0.3 < Z < 1$.
}
\label{fig:Z2}
\end{figure}

For each model, we calculate the $\chi_{\nu}^{2}$\ 
value and a nonparametric measure of the fraction of equivalent widths above or 
below the fit model (Figure ~\ref{fig:Z1}). 
The probability of a certain fraction of observations randomly falling below/above a 
line is given by the binomial theorem.  To obtain an acceptable $\chi_{\nu}^{2}$\, 
we add to the equivalent widths a line-of-sight variation along the
lines of \citet{miller2013}, where we consider the values $\sigma=5.0,7.5$ m\AA . 
At $Z<0.3$, the $\chi_{\nu}^{2}$\ value rises significantly above the best fit and too many
equivalent widths lie above the model (Figure ~\ref{fig:Z1}, ~\ref{fig:Z2}). 
The restrictions at high metallicity are weak, so we assume that the halo gas is 
unlikely to be supersolar, leading to a metallicity range of $0.3-1$ solar,
with a formal best fit value in the middle of that range.
This is in excellent agreement with the values deduced from \citet{faerman16}
and \citet{zhijie2018}.

\clearpage

A separate analysis of the metallicity can be determined in the sightline to the LMC
because we have both an electron column (from the pulsar dispersion measure) and 
a O VII equivalent width \citep{wang05, yao2009, fang13, miller2013, miller2015, 
miller2016a, hk2016}.  In the optically thin limit, this would lead to a metallicity
for the hot gas phase of about 0.3 solar.  When including optical depth effects, one must
include the rotation of the hot halo, 183$\pm$41 km s$^{-1}$ \citep{hk2016} and 
turbulent motion. The inferred metallicity of the gas is always greater than 0.6 solar, 
and for $b \approx 100$ km s$^{-1}$, the best-fit metallicity is solar \citep{miller2016a}. 
For a stationary halo, the metallicity would be about twice solar for the same Doppler $b$ parameter.

\subsection{The Density Models of \citet{nica2016}}

\citet{nica2016} presents an analysis of Milky Way O\ VII
absorption line data, using both high latitude sightlines (extragalactic) and
absorption from low latitude sources, which primarily lie in the disk and
bulge. \ They present a variety of models, among which one (M3) contains the missing
baryons in the Milky Way within a radius of 1.2R$_{virial}$. \ We calculated
the emission measure associated with model M3 at $b=90^{\circ}$ and find it to
exceed the observed values by about an order of magnitude (this also occurs for other directions). 
Model M3 is a combination of an exponential cylindrical model (M2) and a
spherical $\beta$ model, where the first component produces an emission measure
of $6.\,\allowbreak4\times10^{-2}$ pc cm$^{-6}$, and the second component has
an emission measure of $\allowbreak9.\,\allowbreak1\times10^{-2}$ pc cm$^{-6}%
$, or a sum of $\allowbreak1.55\times10^{-1}$ pc cm$^{-6}$ (toward
$b=90^{\circ}$). This is 12 times greater than the high latitude emission measure of
$1.25 \times 10^{-2}$ pc cm$^{-6}$ \citep{mccammon02} used
as a point of comparison. \ This is also true for their Model B, where the
overproduction factor is 14.5 relative to the emission measure of $\allowbreak1.25\times
10^{-2}$ pc cm$^{-6}$.
The overproduction factors are 37, 130, and 18 for models A (spherical $\beta$ model),
M1 (exponential spherical model), and M4 (spherical $\beta$ model), respectively. 
We also compared the predicted O VII and O VIII line strengths, corrected for
optical depth effects \citep{yunyang2017}, with the observed values 
\citep{hs12,hs13} and found a similar discrepancy between their model and the data. 

\begin{figure}       %%%%%%%%%%%%% Figure 6
\centering
%\epsscale{0.8}
\plotone {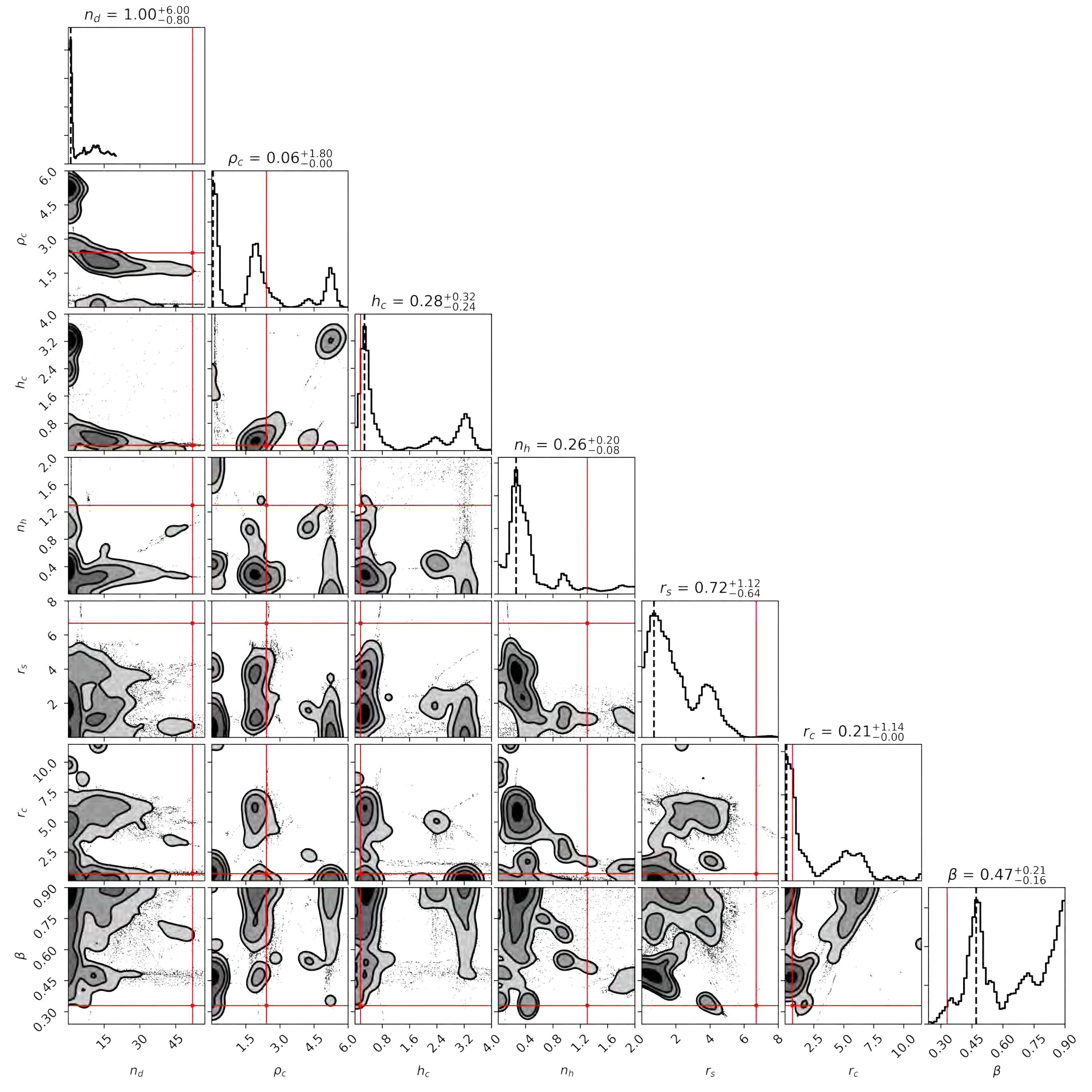}
\caption{
The probability density results of a fit of emission line data to model M3 of \citet{nica2016}, 
which is their high gaseous halo mass model and is a combination of a spherical $\beta$ model
with a central cavity and an exponential cylindrical model. A best-fit is not found and the 
regions of high probability density usually do not correspond to the \citet{nica2016} model values,
shown as the intersection of two red lines. 
}
\label{fig:NicastroM3}
\end{figure}

For their model M3, a combination of models A and M2, they used their best fit
for model M2, then fixed those parameters and added the spherical $\beta$
model (A). They froze $\beta=0.33$ and $R_{c}=0$, although their best-fit
values in the spherical $\beta$ model were $\beta=0.62$ and $R_{c}=5.6$.
This high-mass model does not consider the available parameter space, raising a uniqueness concern, 
so we searched for a self-consistent model. \ We used the emission line data for the fitting, as 
it has many more sightlines and more statistical power than the absorption line data.
We corrected for optical depth effects as described in \citet{yunyang2017} and 
employed a MCMC fitting approach with a sample size of $1.2\times10^{6}$ and 
with the seven free model parameters (Figure ~\ref{fig:NicastroM3}).
We fail to find a global best-fit in that there are often
multiple regions of comparable probability density. \ These regions of higher
probability density usually do not correspond to the parameters adopted in
model M3 of \citet{nica2016}.
We cannot confirm the model of \citet{nica2016}, 
as it significantly overpredicts the emission line observations and because 
we cannot find a self-consistent solution to their favored high-mass model. 

\subsection{Constraints on the Density Distribution from Temperature Measurements}

As discussed above, in hydrostatic equilibrium, $\beta = T_{rot}/T$, where $T$ is the gas
temperature and $T_{rot}$ is from the rotational velocity.  Therefore, measurements of $T$
provide important constraints on the radial density distribution.
One technique for determining the temperature of the Milky Way's hot gas component is 
to measure the O\ VIII to O\ VII absorption line ratio along background quasar sightlines.  
If one assumes the O\ VII and O\ VIII lines originate from the same gas phase, the ratio 
of the column densities is a temperature diagnostic since the ion fractions of these 
species change relative to each other in the expected temperature range of the gas.  
Local O\ VIII absorption is detected less frequently than O\ VII absorption due to signal
to noise limitations. However there are well-known detections of local O\ VII and 
O\ VIII absorption in several quasar spectra including 3C 273, Mrk 421, and PKS 2155 
\citep{rasm2003,williams05,neva2017}.  From these O\ VII and O\ VIII equivalent widths, 
column density ratios can be determined, from which we use standard collisional 
ionization models to infer temperatures of $1.5-2 \times 10^6$ K.

X-ray emission lines are also a useful diagnostic of the Milky Way's hot gas 
temperature and density distribution.  Studies of X-ray emission lines, typically 
the same O\ VII and O\ VIII ions as absorption studies, have varied from single 
observations of a blank field of sky to comprehensive studies of the halo gas using 
X-ray observations covering the entire sky.  For example, \cite{mccammon02} observed 
a 1 sr region of the sky toward $l = 90^{\circ}, b = +60^{\circ}$ using a quantum 
calorimeter sounding rocket and were able to fit the spectrum of the absorbed 
soft X-ray background with a collisional ionzation model with an emission measure 
of $3.7 \times 10^{-3}$ pc cm$^{-6}$ and temperature of $2.6 \times 10^{6}$ K.  
Alternatively, \cite{hs13} measured the hot gas temperature by fitting 110 high-latitude 
\textit{XMM-Newton} observations with collisional ionization plasma models (APEC).
Their spectral fitting results include emission measures ranging from 
$0.4-7 \times 10^{-3}$ pc cm$^{-6}$ and a median temperature 
measurement of $2.2 \times 10^{6} K$ with an interquartile range of $0.63 \times 10^{6}$ K.  
These temperature and emission measure constraints from X-ray emission studies are 
consistent with each other, but differ slightly from absorption line studies.  

\begin{figure}       %%%%%%%%%%%%% Figure 7
\centering
\plotone {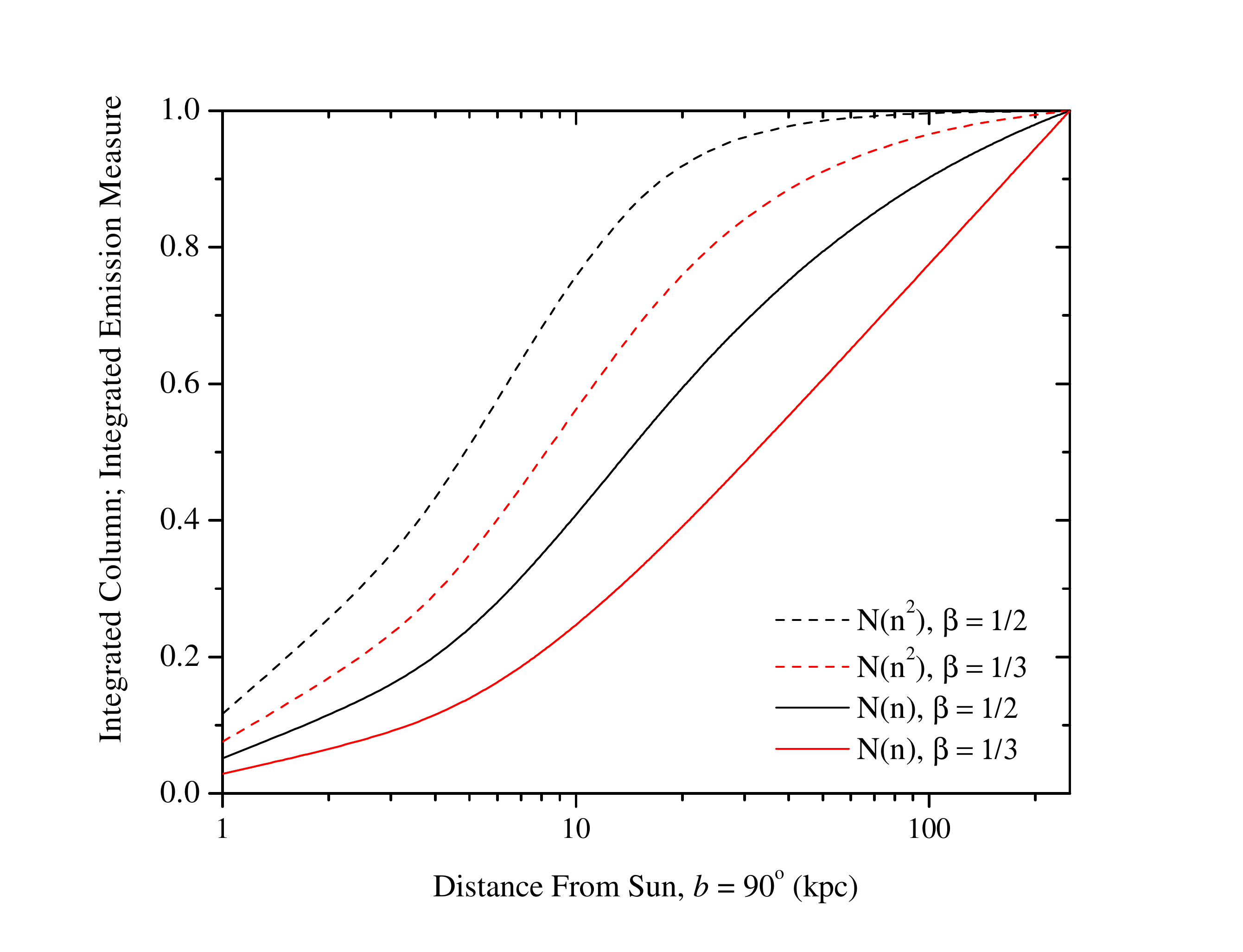}
\caption{
The column density and integrated emission measure, normalized to unity at 
250 kpc, in the direction of $l=90^{\circ}$ for $\beta = 1/3, 1/2$.
For $\beta = 1/2$, 50\% of the emission lies within 4.9 kpc of the Sun and 90\% within 17.6 kpc.
For the column density, 50\% lies within 14.0 kpc and 90\% within 100 kpc. 
With $\beta = 1/3$, the 50\% and 90\% distances are 8.3 kpc and 45.4 kpc for 
the emission measure, and 32.2 kpc and 168 kpc for the column density.  
This shows that Galactic O VII and O VIII absorption and emission studies are
dominated by gas within about 50 kpc.  Also, emission is more weighted by nearer gas than absorption.
}
\label{fig:NEM90}
\end{figure}

The minor discrepancy between the absorption and emission constraints on the halo 
gas temperature may not be significant, but it may indicate a temperature gradient 
to the halo gas.  This is due to emission and absorption measurements weighting 
different parts of the halo since emission processes are proportional to $n^{2}$ while 
absorption processes are proportional to $n$ (Figure ~\ref{fig:NEM90}).  If we assume the denser gas is closer to the 
center or plane of the Milky Way, the larger temperature inferred from the emission 
line measurements is possibly representative of gas closer to the Milky Way as 
opposed to the lower temperature inferred from the absorption line measurements.  
This is not a strong constraint however, and thus we adopt a temperature of $2 \times 10^6$ K as
being representative of the Milky Way hot halo at kpc distances.
This temperature is approximately the virial temperature for the Milky Way.

One can determine the virial temperature from the rotational velocity 
of the Galaxy, taken to be 240 km s$^{-1}$ \citep{bovy2014, reid2014},
$T_{rot} = 1.4 \times 10^6$ K at 20 kpc from the center.  The precise
radius used is unimportant because the rotational velocity changes slowly
for a NFW profile.  The ratio of $T_{rot}$ to the observed halo temperature
is approximately $T_{rot}/T \approx 0.7 = \beta_{spec}$, 
which is a bit steeper than the value inferred from the X-ray line
studies. The difference may be attributable to turbulent motion providing additional support against gravity. 
The level of turbulent support that would bring $\beta_{spec} \approx 0.5$ is
a gas where the turbulent Mach number is about 0.5, which can occur in simulations \citep{fielding16}.

We consider whether it is possible to have density profiles that are 
significantly flatter than $\beta = 0.5$, as this has important implications
for the gaseous mass of the hot halo.
In one of the flatter density distributions, \citet{feld2013} have a temperature 
that rises to about $7\times10^{6}$ K at 2 kpc from the midplane, decreasing 
to $4\times10^{6}$ K at 10 kpc and $3\times10^{6}$ K at 20 kpc (Feldmann, private communication).
Another flatter profile is given in the model of \citet{kauf08}, where the halo 
is hot enough that the density has a radial dependence of about $n \propto r^{-1}$ 
for $15 < r < 50$ kpc. For a hydrostatic model and a Milky Way potential, this 
would correspond to a temperature of about $4T_{rot} \approx 6 \times 10^6$ K. 

\begin{figure}       %%%%%%%%%%%%% Figure 8
\centering
\plotone {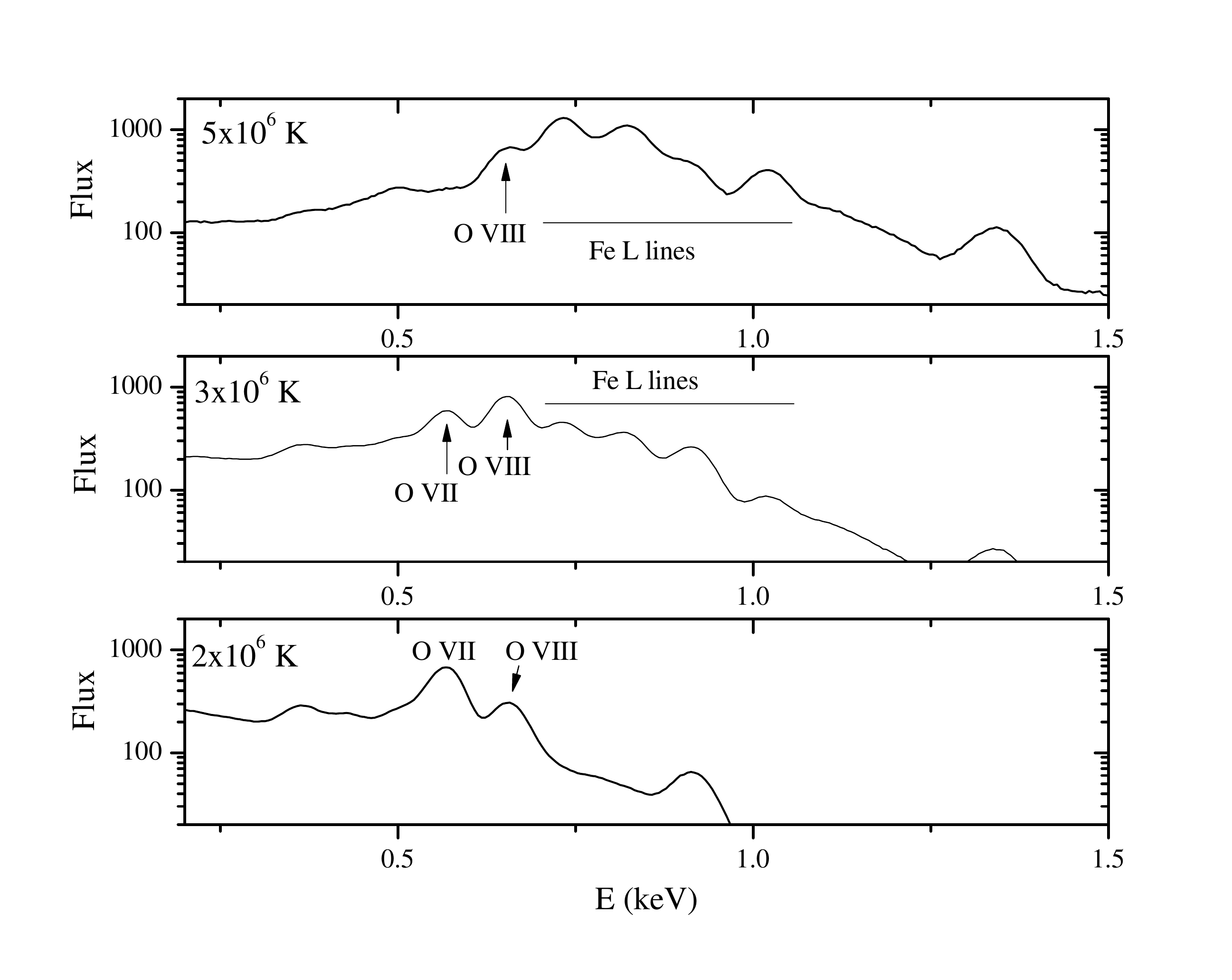}
\caption{
Model spectral energy distributions (count rates) at XMM resolution shows dramatic
changes as a function of temperature. Except for special directions (e.g., SNR), 
observations resemble the bottom panel, a $2\times10^{6}$ K plasma and exclude significant amounts of gas at higher temperatures within $\sim$30 kpc of the galaxy.
}
\label{fig:XMM3spec}
\end{figure}

As the X-ray emission is dominated by material within 20 kpc of the disk (see below), due
to the density squared dependence of the emission measure \citep{miller2015,hk2016}, 
these high temperatures would have a striking spectral energy signature (Figure ~\ref{fig:XMM3spec}).
For the preferred halo temperature of about $2\times10^{6}$ K, 
the O VII emission is stronger than the O VIII emission. 
This relative line strength ratio would be reversed by $3\times10^{6}$ K, 
and at $5\times10^{6}$ K the O\ VII line is not longer detectable while 
the Fe L complex becomes quite prominent. The general lack of the spectral
energy signatures of higher temperature gas \citep{mccammon02, hs13} argues
against the models of \citet{feld2013} and \citet{kauf08}.

\subsection{The Milky Way Gaseous Halo Mass Inferred From The Density Distributions}

\begin{figure}       %%%%%%%%%%%%% Figure 9
\centering
\plotone {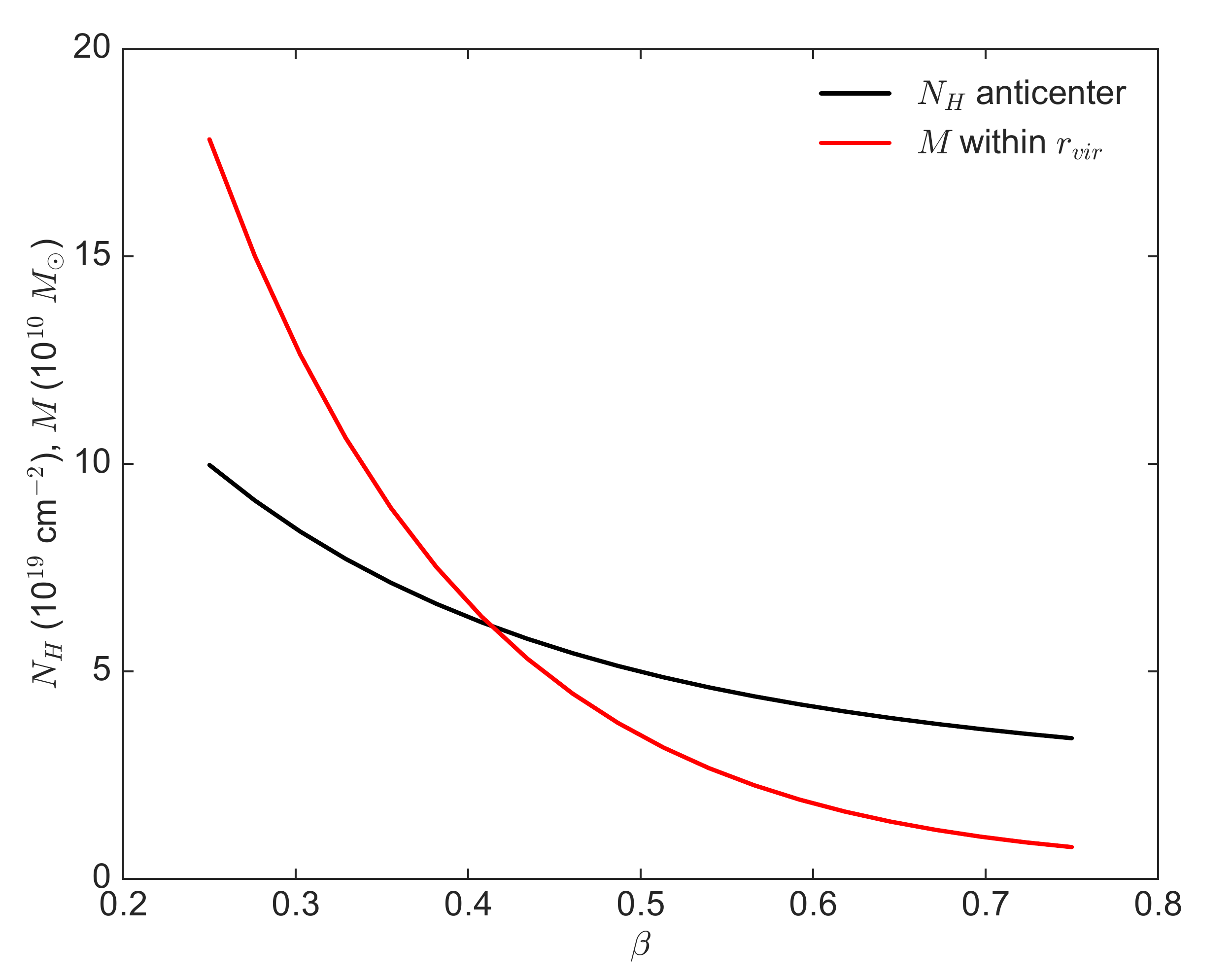}
\caption{The hydrogen column density and gaseous mass from 10 kpc to the virial radius
of 250 kpc. The models have a power-law density distribution, given by $\beta$, and all
models are constrained to reproduce the pulsar dispersion measure to the LMC.
For $\beta \approx 0.5$ \citep{miller2015}, the hot halo accounts for only about one-third
of the missing baryons. In very flat density distributions, $\beta \lesssim 0.25$, the hot halo
could account for the missing baryons, which lies in the range $8-24 \times 10^{10}$ $M_{\odot}$.
}
\label{fig:NhMbeta}
\end{figure}

The mass of hot gas in the halo depends on the density distribution with
radius, which is constrained in that we know the electron column density
toward the LMC \citep{ander2010}. The density distribution is given
by a normalization and, for a power-law distribution, a power-law index.
Without further constraints, a degeneracy exists between the power-law density
slope and the density normalization in the sense that the same electron column
is obtained with a lower normalization and a flatter power-law distribution or
a higher normalization and a steeper power-law distribution. The flatter
density distributions lead to larger gas masses when the density law is
extrapolated to the virial radius. As an example, for a density law of
$n=n_{0}(r/r_{0})^{-3\beta}$, the mass within 250 kpc is an order of
magnitude higher for $\beta = 0.2$ than for $\beta = 0.6$ (Figure ~\ref{fig:NhMbeta}).

This degeneracy is the source of the controversy between authors for the gas
mass contained within the virial radius of the Milky Way. \ When a fairly flat
density law is considered ($\beta\approx0.2-0.3$; \citealt{gupta12, gupta14,faerman16}),
the hot halo can contain the missing baryons, whereas for the models discussed
by \citet{miller2015} and \citet{yunyang2017}, with $\beta\approx0.5$, the gaseous halo mass is
significantly less and does not account for the missing baryons 
(about half are missing atR$_{200}$). This is seen
in Figure ~\ref{fig:NhMbeta}, where the electron density is constrained to reproduce
the pulsar dispersion measure, while we calculate the mass within 250 kpc and
the electron column radially outward from the Sun to 250 kpc. At the lowest
values of $\beta$ ($<0.3$), the gaseous halo mass rises into the range of
the missing baryons, which has a value in the range $0.8-2.4\times10^{11}$ M$_{\odot
}$, depending on the assumed total mass for the Milky Way ($1-2\times10^{12}$
M$_{\odot}$; \citealt{xue2008, gnedin2010, watkins10, barber14, bland16}).

The challenge is to determine the power-law density index from other
observations and there are a few approaches to this problem. \ One method is
to use information inferred from ram-pressure stripping of dwarf galaxies or
from the interaction of the Magellanic Clouds and the Magellanic Stream with
the ambient hot halo. \ These require extensive hydrodynamic modeling and
knowledge of the orbit of the object, along with the assumption that ram
pressure is the primary physical process. 
The most thorough examination of this problem is by \citet{emerick16}, who considered 
both ram pressure stripping as well as feedback effects from stimulated star formation.
They find that these processes are unable to account for stripping (quenching) in less
than 2 Gyr and discuss additional physics that might be included. 
This indicates that there is not a good understanding of gas loss from Local Group
dwarf galaxies, so using them to infer the ambient density may be problematic, leading
to significant uncertainties. 

A different and promising approach to inferring the ambient density comes from 
a recent work that models the ram pressure on the leading edge of the LMC gas disk \citep{salem15}.
By fitting their model to the detailed information available, 
they produce a particularly good density determination.
\citet{salem15} give a gas density of $1.1_{-0.45}^{+0.44}\times10^{-4}$ cm$^{-3}$
at R$\approx50$ kpc. 

\begin{figure}       %%%%%%%%%%%%% Figure 10
\centering
\plotone {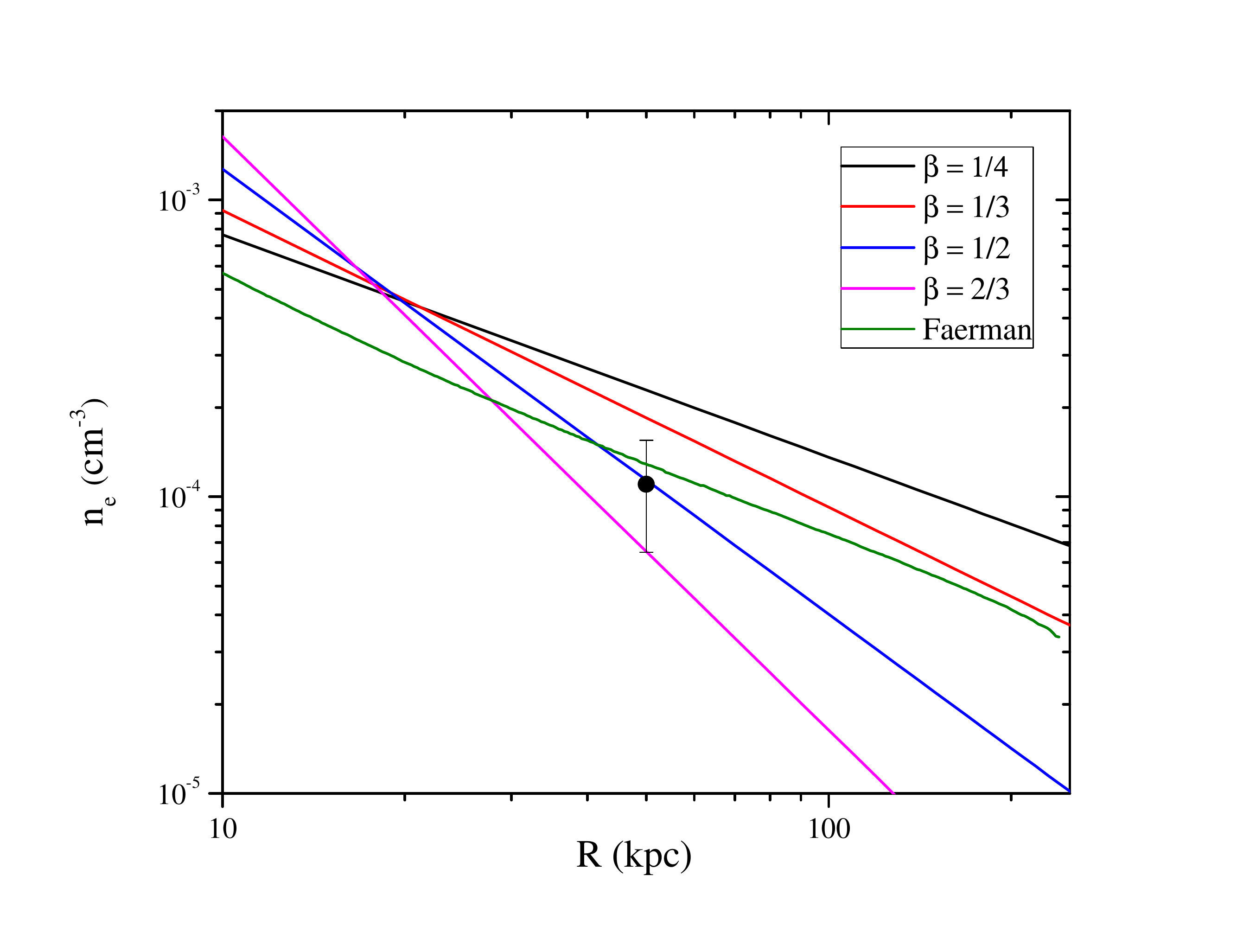}
\caption{
Power-law density distributions plus the density distribution from \citet{faerman16}, 
normalized to reproduce the pulsar dispersion measure toward the LMC. The data point,
from the numerical modeling of the LMC gas \citep{salem15}, is consistent with all but 
the flattest density distributions.
}
\label{fig:salem}
\end{figure}

Density constraints were also deduced in a model where there is a shock cascade between the 
Magellanic HI Stream and the ambient hot halo medium \citep{tepp2015}.  
There are uncertainties in the inferred ambient density of a factor of two, plus the distance to the Stream 
may be larger than adopted, leading to further uncertainties in the hot ambient density.  
Therefore, we use the ambient halo density from \citet{salem15}, which appears more secure. 
We compare the \citet{salem15} result to various density laws that are
already constrained to reproduce the electron integral to the LMC\ (pulsar
dispersion column) and find it to be consistent with all but the flattest of
density profiles (Figure ~\ref{fig:salem}), requiring $\beta \gtrsim 0.3$.

To summarize, the observational data for the Milky Way points to $\beta
\approx0.5$, and after corrections for optical depth \citep{yunyang2017}, 
the hot gaseous mass of the hot halo is M($< 250$ kpc) = $2.8 \pm 0.5 \times 10^{10}$ M$_\odot$, 
similar to prior values \citep{miller2015}, and the exponential hot disk mass is 
about 1\% of the halo mass, with a value of $1.8 \pm 0.3 \times 10^8$ M$_\odot$. 
This hot halo gas mass is less than or comparable to the stellar mass of the 
Galaxy and fails to account for the missing baryons by a factor of two \citep{miller2015,salem15,yunyang2017}. 
A caveat here is that these tracers
of the halo gas density are dominated by gas within 50 kpc, so if the gas
density were to flatten beyond 50 kpc, a larger halo mass would occur. \ Such
a flattening of the gas density at larger radii is suggested by \citet{faerman16},
who show that with such a model, the missing baryons lie within $R_{200}$. 
The hot gas component in their model is not characterized by a single value of
$\beta$, but decreases from 0.35 (8 kpc) to 0.26 (70 kpc), slowly rising to 
0.30 (180 kpc) and then rising up to 0.41 near the virial radius.
The hot gas mass differences between models points out the necessity to constrain the shape of the
density law in the range $0.2R_{200}-R_{200}$ (50--250 kpc).

\begin{figure}       %%%%%%%%%%%%% Figure 11
\centering
\plotone {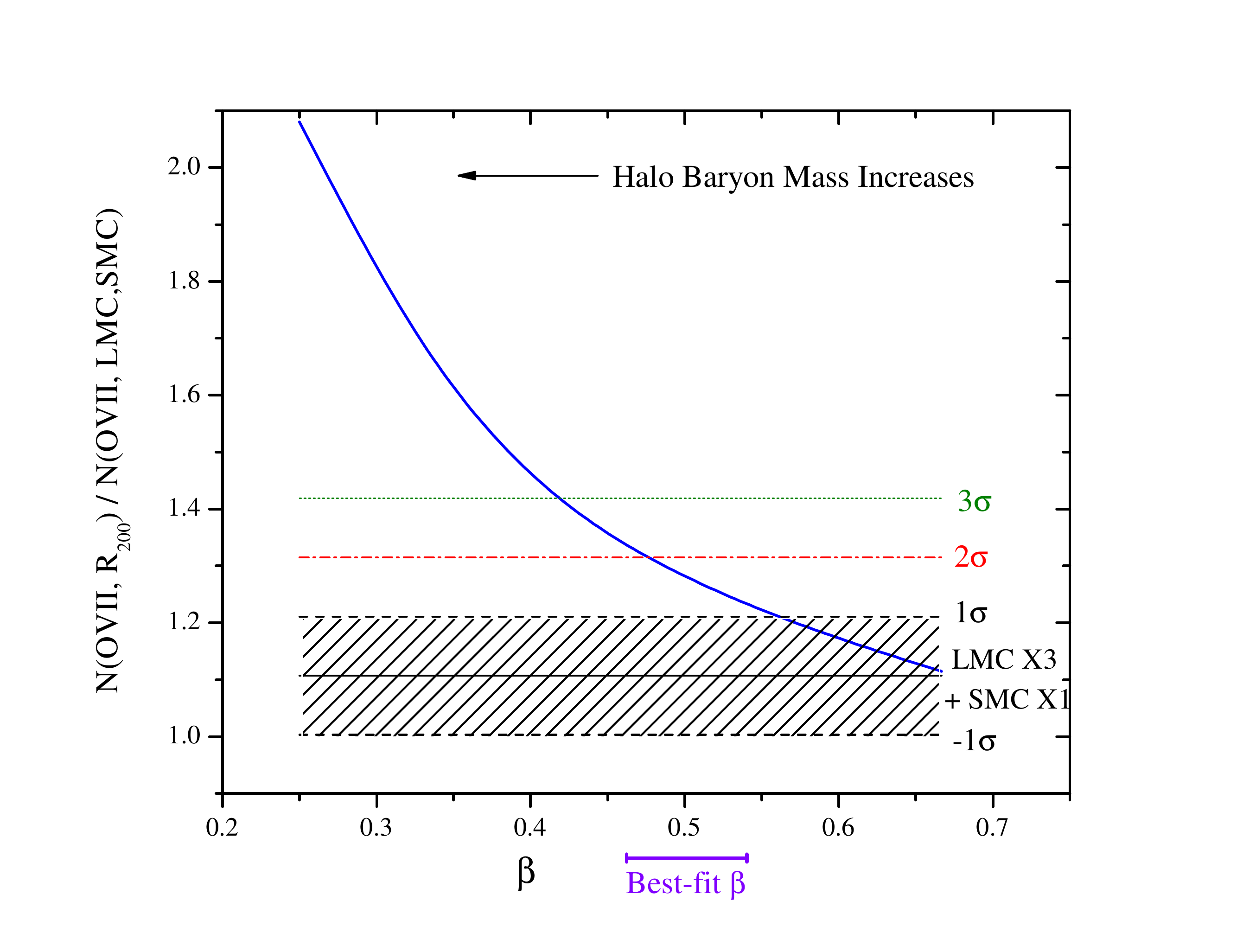}
\caption{
The ratio of the O VII column to $R_{200}$, from \citet{miller2015} to the weighted average O VII 
column measured toward LMC X-3 and SMC X-1.  As the two columns are nearly 
equal, a significant fraction of the column does not lie beyond the Magellanic Clouds. 
At the 3$\sigma$ level, this implies that $\beta > 0.42$.
}
\label{fig:LMCX3SMCX1}
\end{figure}

One of the few constraints for the density in this range comes from the
realization that for the flatter density profiles ($\beta<0.4$), there is a
significant contribution to the column density in the 50--250 kpc range. We can
obtain the column for $R<50$ kpc from the observation toward LMC\ X-3 and SMC X-1, for
which we measure the observed O\ VII equivalent widths from archival \textit{XMM-Newton}
data and obtain values of $23.1 \pm 3.0$ m\AA\ and $21.0 \pm 4.9$ m\AA . 
The weighted mean of these two sightlines is $22.5 \pm 2.6$ m\AA .
We can compare this to the equivalent width inferred from background AGN, which samples the 
entire halo out to and beyond $R_{200}$. Those observations toward AGNs,
represented by the model of \citep{miller2015}, imply that the equivalent
width through the halo in that direction is 24.9 m\AA . 
As this is nearly the same as the value toward the LMC/SMC objects, it limits
the amount of the O VII column that lies beyond. \ To quantify this, we
calculate the ratio of the column within 250 kpc to that within 50 kpc in the
direction of the Magellanic Clouds, as shown in Figure ~\ref{fig:LMCX3SMCX1}. 
We find $\beta > 0.42$ at the 3$\sigma$ level and $\beta>0.34$ at the 5$\sigma$ level,
based on the weighted mean EW and the best-fit model EW to 250 kpc. 
This would rule out particularly flat density profiles, such as those 
of \citet{gupta12} or \citet{faerman16}.  However, there are two caveats.
In addition to the statistical uncertainties used, there can be significant line of sight 
variations in the absorption column \citep{miller2013}, which we estimate to be 
22\% based on the variation of groupings of sightlines at similar Galactic latitude and
longitude. This concern could be addressed by using several lines of sight 
toward the LMC and SMC, but such data is not available currently.  
Also, the above analysis assumes a constant oxygen abundance to 250 kpc,
while a significant decline in the oxygen abundance beyond 50 kpc would lead
to only a modest increase in the O VII equivalent width even with a $\beta < 0.3$
profile. 

To conclude, we do not find compelling evidence for a flattening of the halo
gas density in the range 50--250 kpc, although further observations are needed to gain
insight into this important issue. Without such a flattening, the gas 
within $R_{200}$ fails to account for the missing baryons by a significant margin.

\subsection{External Galaxies}

The purpose of considering external galaxies to to determine whether their 
hot halo properties agree or disagree with the insights obtained from Milky
Way studies. 
From X-ray studies, one obtains the surface brightness distribution, which can
be converted to a density profile when a temperature and a metallicity are known.
Temperature fitting requires more photons than obtaining a surface brightness
distribution while metallicity fitting requires about an order of magnitude more
photons, with current X-ray instrumentation.

For the well-studied case of early-type galaxies, the temperature of the hot gas exceeds
$T_{rot}$, typically by a factor of 1.4--2 \citep{davis96,loew99,david2006,athey2007, 
pell2011, goulding16}, presumably reflecting the additional heating of the gas 
from supernovae and AGN (e.g., \citealt{gaspari14}).  Observers also
find that the temperature gradient is smaller than the density gradient, so for
the isothermal case, this leads to $dln(n)/dln(r) = -3T_{rot}/T$
(equivalent to $\beta = 0.5-0.7$ for the typical temperature ratio range; 
\citealt{goulding16} give a median value of $\beta_{spec} = 0.6$).  

This is similar to the density decrease that is inferred from the surface 
brightness decline, so there is consistency between the two methods.  
Where $\beta$ derived from surface brightness measurements is smaller
than $T/T_{rot}$, this may be an indicator that turbulent motion is an important component
of the pressure support (e.g., \citealt{fielding16}).

\begin{figure}       %%%%%%%%%%%%% Figure 12
\centering
\plotone {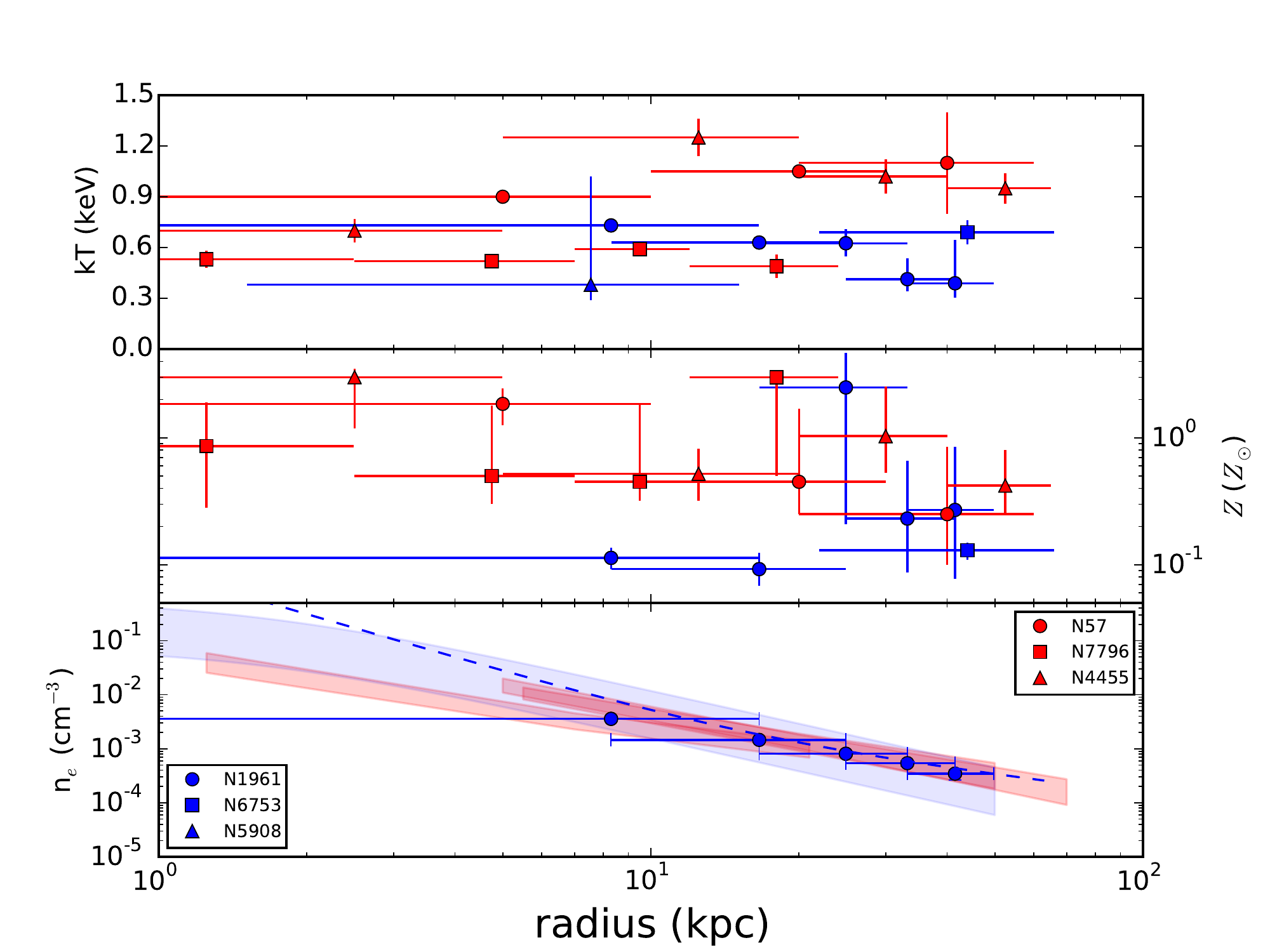}
\caption{
The temperature, metallicity, and electron distribution for three massive spirals (blue),
two elliptical galaxy (NGC 57, NGC 7796, in red) and a massive early-type galaxy (NGC 4455, red);
all are isolated systems. The temperature distribution is often isothermal, although there is a 
rise of T in NGC 4455 relative to the center and a decrease in T for NGC 1961. The metallicities
of the early type systems are generally larger than those of the late-type galaxies, with a median
value of about 0.3 solar at 30 kpc. The density distribution is similar for all galaxies (red-pink 
bands are for early type galaxies and blue are late-type galaxies; 
the dashed blue line is a $\beta = 0.5$ model.
}
\label{fig:anderson6gal}
\end{figure}

Figure ~\ref{fig:anderson6gal} shows the temperature, metallicity, and density 
profiles measured for three isolated ellipticals and three isolated spirals. 
This figure is adapted from Figure 14 of \citet{ander2016}, with two 
new spiral galaxies added to improve the comparison. These are all galaxies more massive 
than the Milky Way ($M_* > 10^{11} M_{\odot}$) and have specifically been selected not 
to lie in larger galaxy cluster or group environments, so there is no contamination
from an intracluster or intragroup medium. 

In the temperature profiles, there is no clear difference between spiral and 
elliptical galaxies. At large radii ($r \gtrsim 20$ kpc), where the interstellar 
medium of the galaxy is no longer contributing to the observed signal, there is 
some suggestion that the spiral galaxies have cooler hot haloes than the elliptical galaxies. 
Spiral galaxies inhabit lower-mass halos than elliptical galaxies at fixed stellar mass, 
so the observed result could reflect hydrostatic equilibrium with a lower-mass halo 
for spiral galaxies (see also the discussion in \citealt{ander2016}). 

The metallicity profiles show a clear difference between non-starburst spiral and elliptical 
galaxies. The hot halos of the spiral galaxies NGC 1961 and NGC 6753 are clearly 
sub-solar, at $0.1-0.2 Z_{\odot}$, while the hot halos of the elliptical galaxies 
are close to Solar abundance or even super-Solar. 

The radial density distributions of the hot halos around isolated non-starburst
spiral and elliptical galaxies are fairly similar.  In general, the profiles are 
consistent with $n \propto r^{-3/2}$ (or $\beta = 0.5$), which is obtained over the distance 
range 10--100 kpc \citep{humph06, ander2011, Dai2012, ander2013}.  This is also
similar to the hot gas density distribution that is found in the optical regions
of early type galaxies (e.g., \citealt{athey2007}) but is a flatter distribution 
than found in clusters of galaxies (e.g., \citealt{vikh06}). 

For the individual massive spiral galaxies, NGC 1961 and UGC 12591, 
$\beta = 0.47 \pm 0.06, 0.48_{-0.08}^{+0.25}$ and $T = 8.0 \pm 0.8\times10^6 $ K, 
$8.0 \pm 0.5\times10^6 $K, respectively \citep{ander2011, Dai2012}.  
\citet{bogdan13} examined the  giant spiral, NGC 6753, and found a density profile 
corresponding to $\beta \approx 0.47$, similar to the other galaxies. 

In addition to the three ellipticals shown in Figure ~\ref{fig:anderson6gal}, 
there are also observations of the isolated ellipticals NGC 720 and NGC 1521. 
For the extended emission around the isolated elliptical NGC 720, \citet{humph06} 
fit a profile that at large radii approximates to $\beta \approx 0.46$ and 
$T = 6.0 \pm 0.5 \times10^6 $ K (1--20 kpc). \citet{humph11} find a flatter profile, 
with $\beta \approx 0.41$. \citet{anderson2014} show that the \citet{humph11} 
profile, which is based on a spectral analysis, overpredicts the observed soft X-ray 
surface brightness at all radii. The spatial analysis of \citet{anderson2014} finds $\beta = 0.52 \pm 0.03$, 
and is consistent with the more modest hot halo found by \citet{humph06}.
For NGC 1521, the slope of their best-fit profile corresponds to $\beta \approx 0.44$ 
and $T = 6.0\times10^6$ K. 

These results were obtained from deep studies of individual objects, but we find 
similar results for stacked observations of large populations. 
The radial profile from stacking hundreds of nearby isolated early-type galaxies 
yields $\beta = 0.6 \pm 0.15$ and a similar stack of isolated late-type galaxies 
yields  $\beta = 0.55 \pm 0.1$ (L $\ge$ L*; \citealt{ander2013}, see also \citealt{ander2015}).  

Finally, we note that none of these studies extend to a 
significant fraction of the virial radius, so the total mass within the
virial radius relies on an extrapolation. However, in every case there is 
no evidence for a flattening of the slope at large radius, which would 
be necessary in order for the extrapolations to significantly under-predict 
the total mass in the hot phase. Instead, \citet{ander2016} find that the 
X-ray surface brightness attributable to the hot gas shows a tentative steepening 
beyond about 20 kpc.  In an analysis of NGC 720, there is also a tentative 
steeping at $r \gtrsim 25$ kpc \citep{anderson2014}

In Table 1 we summarize these observations of hot gaseous halos around isolated 
massive galaxies. In general, the trend is that more massive galaxies have more 
mass within 50 kpc and more mass inferred within the virial radius. 
It is unclear if hot halos disappear below L* or if it just 
becomes too faint to be detectable. There also seems to be a hint of a trend 
such that elliptical galaxies have more hot gas mass than spiral galaxies, 
but there are not enough data points to to make definitive conclusions. 

A recent survey obtained XMM observations of five additional galaxies, which together with UGC 12591, constitutes a complete sample of massive spiral galaxies in the local universe (D $<$ 100 Mpc; the CGM-MASS sample; \citealt{jiangtao16,jiangtao17,jiangtao18}); the median value of M$_* = 2.5 \times 10^{11}$ M$_{\odot}$ and the median rotation velocity is about 330 km s$^{-1}$.  
This sample has a range in L$_X$ of about three, for galaxies with similar values of M$_*$, which is less than the factor of 30 range seen in lower mass galaxies \citep{osull2004}.
The temperatures are in the 0.7--1.1 keV range and the density distributions have a range of $\beta = 0.35 - 0.68$, with a median of $\beta = 0.45$.  There is a relationship between the measured radial density distribution ($\beta$) and the ratio of the temperature inferred from rotation to the thermal temperature ($\beta_{spec}$), as physical arguments would predict.  That is, systems that are hotter, relative to their rotational temperature, have flatter density distributions.  

Using median values, $M_{200} = 8 \times 10^{12}$ M$_{\odot}$, so M$_{baryon} = 1.3 \times 10^{12}$ M$_{\odot}$. The median stellar mass is $2.5 \times 10^{11}$ M$_{\odot}$ and the hot gaseous mass is slightly lower, at about $1.5 \times 10^{11}$ M$_{\odot}$, so when including cooler disk gas, the measured baryon mass within $R_{200}$ is $\approx 4 \times 10^{11}$ M$_{\odot}$, or about 41\% of the baryons.  The remainder is unaccounted for and presumably lies beyond $R_{200}$ or in a cooler phase, which we explore in the next section.

\section{The Mass Contributions from UV Detected Absorbing Gas}

Neutral and warm ionized gas is widely detected through UV absorption lines, where the interpretation of the absorption can infer a significant baryon mass (e.g., \citealt{werk2014,proc2017}).  This gas is at $\sim 10^4$ K ($<< T_{vir}$), so it is not buoyant, and if it is not supported by rotation, it would naturally sink into the galaxy at a rate of $M(H)/t_{ff}$, where M(H) is the mass inferred in the halo and $t_{ff}$ is the free-fall time, about 10$^9$ yr. As the mass of warm halo gas has been suggested to be most of the missing baryons, about 10$^{11}$ M$_{\odot}$ for an L* galaxy \citep{werk2014}, the accretion rate would be $\sim 10^2$ M$_{\odot}$ yr$^{-1}$, which is far in excess of the observed accretion rates of such galaxies, generally less than $\sim 1$ M$_{\odot}$ yr$^{-1}$ \citep{leitner11}.  For this reason, we consider whether having a large warm baryonic mass in the halo is the only feasible interpretation.

There have been several works that examine the absorption properties of the region around galaxies.  
We consider the samples where a galaxy-AGN pairing is established before the spectroscopic observation, 
and where the galaxy is not selected based on gaseous properties. 
Also, we consider low redshift systems ($z \lesssim 0.2$) and avoid samples devoted only to dwarf galaxy 
absorption, as they are significantly lower mass than the X-ray emitting galaxies, which are closer to L*. 
Of the samples considered, both \citet{bowen02} and the targeted sample of \citet{stocke13} 
and \citet{keen17} used relatively local galaxies ($z < 0.02$) and obtain HI columns around 
galaxies with luminosities in the range $10^{-2.5}-10^{0.2}$ L*. 
The other sample considered is the COS-Halos program, which used SDSS data to obtain 
galaxy-AGN pairings, for galaxies near and above L*, and at a typical redshift of 0.2 \citep{werk2014, proc2017}. \

\subsection{HI Equivalent Width Distributions}

We compare the samples of COS-Halos to that of the combination of the \citet{bowen02} and the targeted absorption systems in \citet{stocke13} and \citet{keen17} (henceforth, the Stocke-Bowen sample) and find them different in important ways. The first comparison is of the equivalent widths for the two samples, which are treated differently in the investigations.  Multiple components in a single sight line are added together in the COS-Halos sample. There are usually multiple components for a sight line, as judged by low ionization metal lines, but these components often blend together in the higher optical depth HI lines. Thus, separating such lines into components can be model-dependent, especially at higher column densities. Stocke-Bowen identify individual components associated with a single galaxy, so we combine the individual components in order to make a comparison with the values from COS-Halos. Where the values were not listed, we extracted the equivalent widths. Both samples are for systems with impact parameters less than 150 kpc.

\begin{figure}       %%%%%%%%%%%%% Figure 13 (was 10)
\centering
%\epsscale{0.8}
\plotone {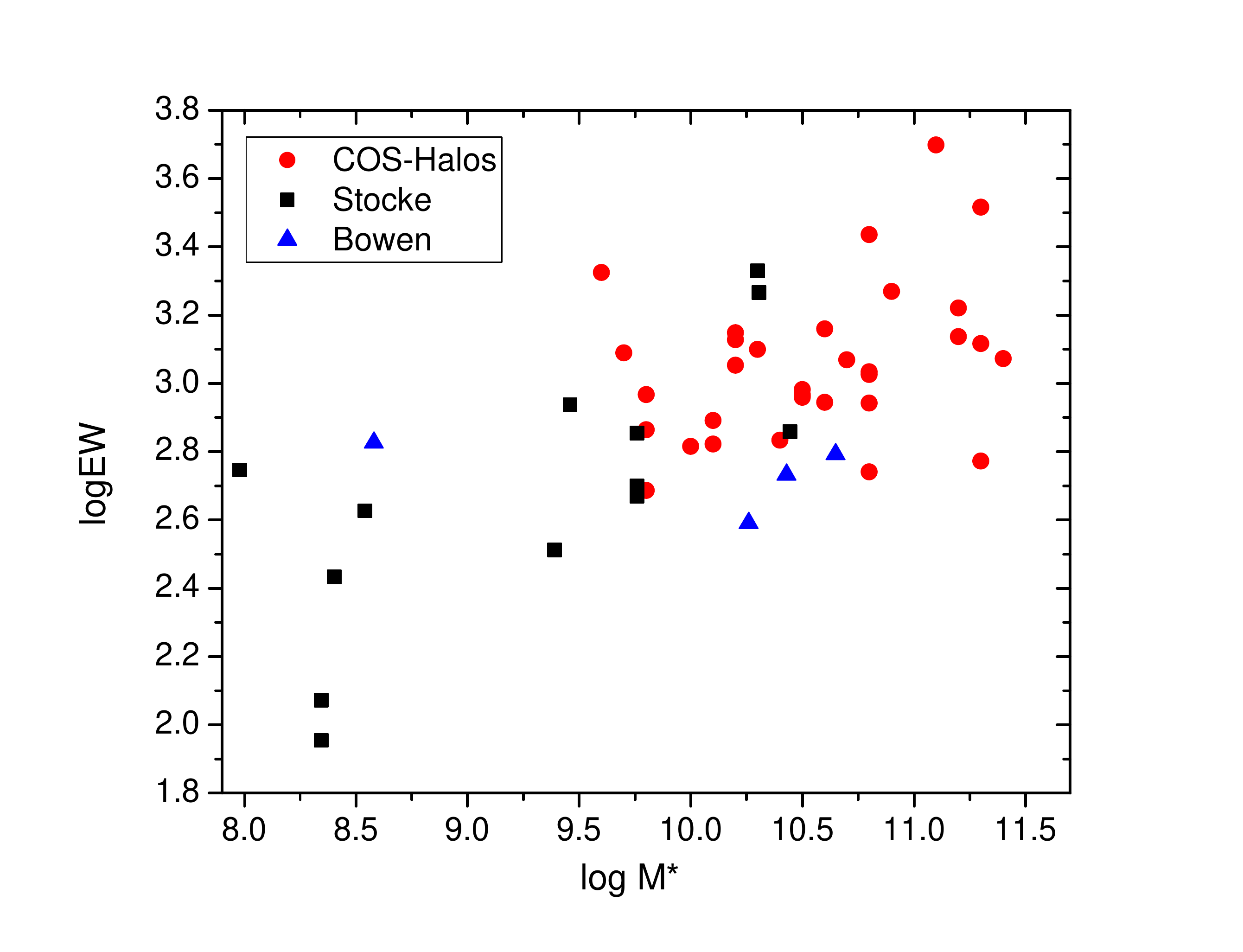}
\caption{
The distribution of HI Ly $\alpha$ equivalent widths from the \citet{bowen02}, from COS-Halos 
\citep{werk2014, proc2017}, and from the targeted sample of \citet{stocke13} and \citet{keen17}.
In the overlap region ($9.4 <$ log M$_*$ $< 10.7$), the equivalent widths the COS-Halos absorption
systems are generally larger than the Stocke-Bowen sample.
}
\label{fig:EWMstarWSB}
\end{figure}

\begin{figure}       %%%%%%%%%%%%% Figure 14 (was 11)
\centering
%\epsscale{0.8}
\plotone {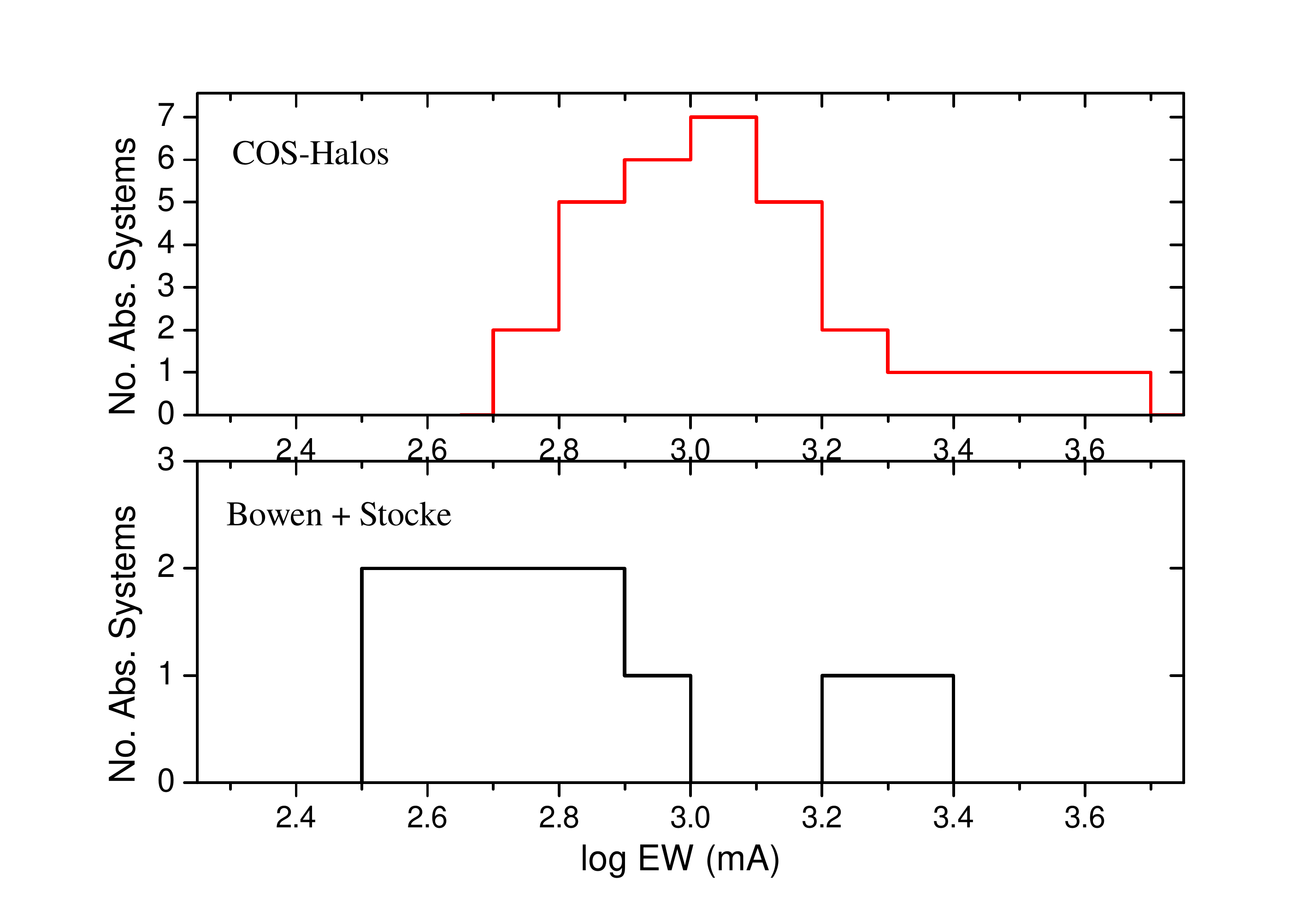}
\caption{
The distribution of equivalent widths for the COS-Halos and the Bowen+Stocke samples
for R $<$ 150 kpc and logM$_* >$ 9.4.  We interpret this as a high equivalent width extension 
plus a distribution of lower equivalent width systems, which could be consistent with separate
disk and halo populations of clouds.
}
\label{fig:HistogramCOSSB}
\end{figure}

The distribution of the equivalent widths is significantly different between the local sample and the COS-Halos sample (Figure ~\ref{fig:EWMstarWSB}).  Part of the difference is because the Stocke-Bowen sample has a number of galaxies with lower luminosities than the lowest luminosity galaxy in the COS-Halos sample. This is significant because there is a correlation between galaxy luminosity and equivalent width. When we just consider galaxies with log$M_*$ $> 9.4$, there is still a difference between the two samples as seen in Figure ~\ref{fig:HistogramCOSSB}. In the Stocke-Bowen sample, there appears to be two systems with equivalent widths above 1800 m\AA, while the rest are more than a factor of two lower (90--864 m\AA).  Below we argue that these are separate populations and the lower equivalent width sample has a median of 540 m\AA.  The COS-Halos sample also has a set of large equivalent width systems, so when we exclude the four with the largest values, the median equivalent width is about 1100 m\AA\, about a factor of two larger than the Stocke-Bowen sample. 
A factor of two difference in a saturated line corresponds to about a factor 
of 50 in the column density, since the equivalent width is proportional to $ln(N^{1/2})$.  
This assumes that the distribution of Doppler $b$ parameters is the same in the two samples. 

%\clearpage

For the highest equivalent widths, there is a dependence on radial distance and stellar mass in the Stocke sample.  Both high equivalent width systems occur in the their most optically luminous galaxies (Figure ~\ref{fig:EWMstarWSB}) and at relatively close radii ($r \leq$ 53 kpc, whereas the sample median is 65 kpc).  The Bowen sample does not have high equivalent width systems, but in the COS-Halos survey the highest equivalent widths tend to occur among the more luminous galaxies (Figure ~\ref{fig:EWMstarWSB}). 

\subsection{The HI Column Density Distributions}

The conversion of equivalent widths to column densities can be uncertain due to the presence of optical depth effects at moderate opacities that define the “flat” part of the curve of growth. This problem can be overcome if there are several lines of different opacity, permitting a curve of growth fit to be achieved. The situation is more challenging for a single saturated line, for which a lower limit can be assigned if there is no knowledge of the Doppler b parameter or of multiple components.  However, if there are other low ionization lines of lower opacity, the Doppler b parameter can be constrained and that information is used in estimating a column from an equivalent width.  Such information was brought to bear on the COS-Halos survey, in which they have obtained best-estimates for N(HI).  For the Stocke sample, Voigt profiles were fitted to the Ly $\alpha$ absorption lines, allowing for multiple components, procedures fully explained by \citet{keen17}. \citet{bowen02} also fit Voigt profiles and assessed the uncertainties through simulations. 

\begin{figure}       %%%%%%%%%%%%% Figure 15 (was 13)
\centering
\plotone {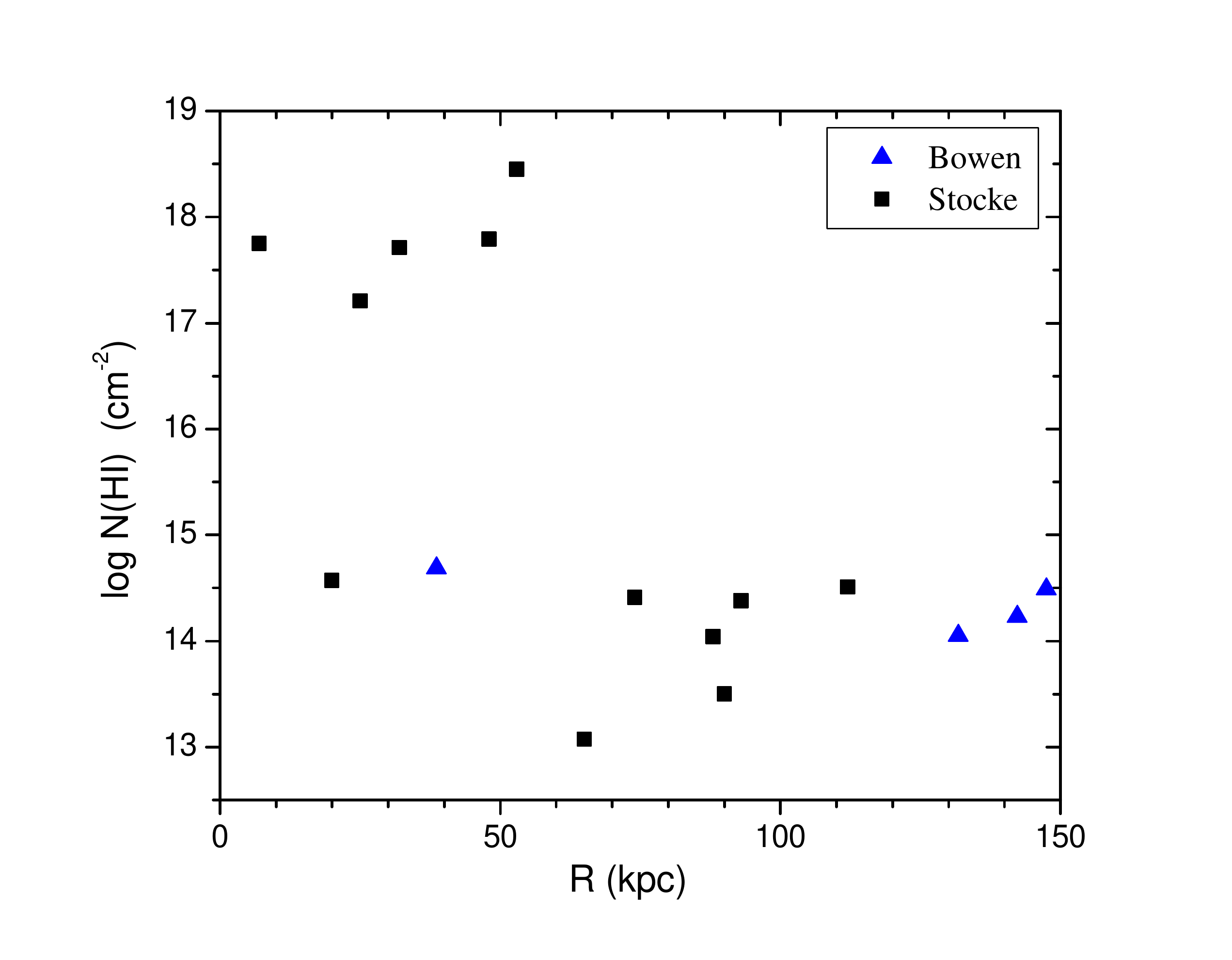} 
\caption{
The HI column determined in the Stocke-Bowen sample as a function of impact parameter for the full range of $M_*$ and for R$<$ 150 kpc.
The high column density population all occur for R $\leq$ 53 kpc may represent
a disk component while the lower column, more extended absorbers may be a halo component.
}
\label{fig:NHIRSB}
\end{figure}

\begin{figure}       %%%%%%%%%%%%% Figure 16 (was 15, 16)
\centering
\plotone {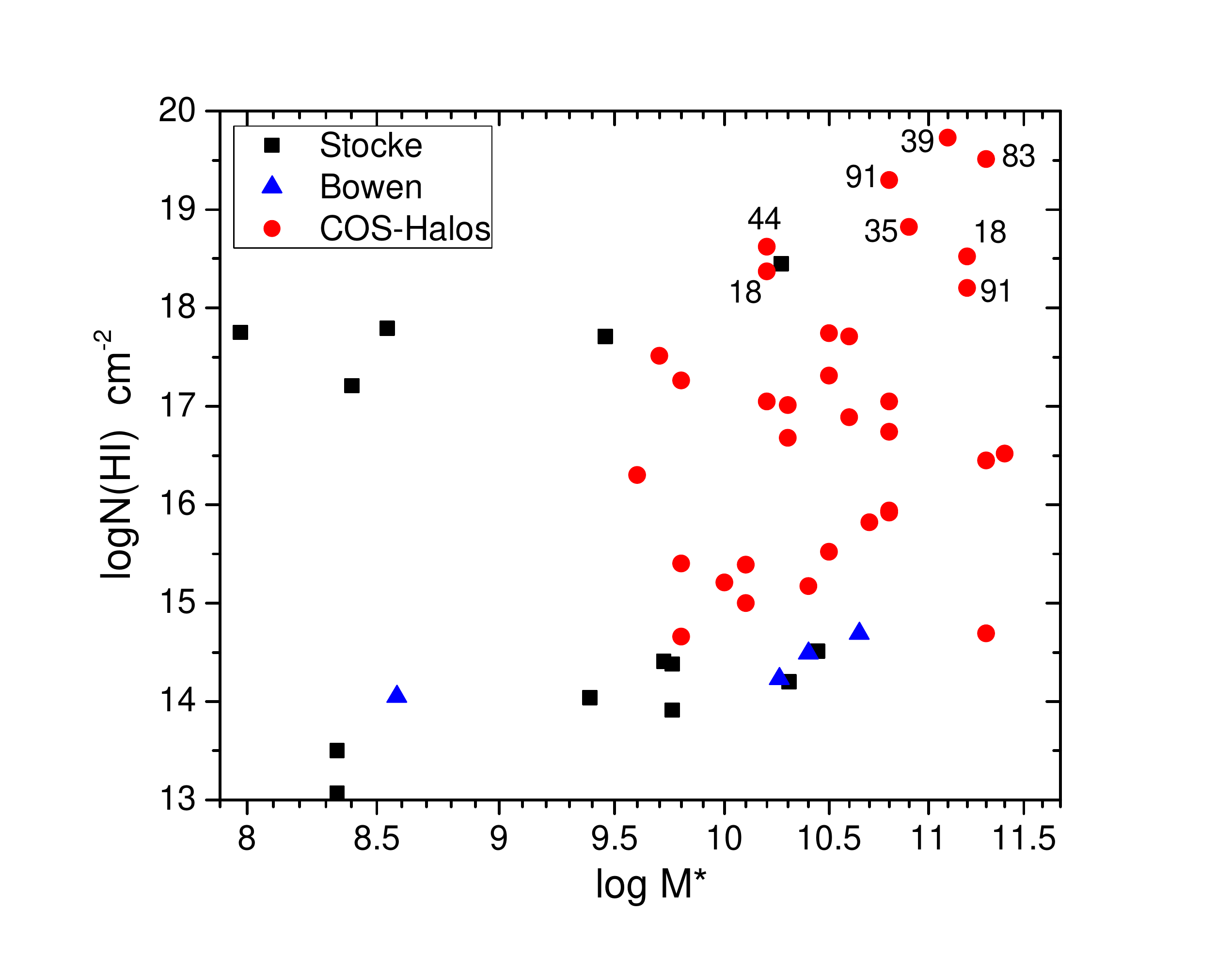} 
\caption{
The HI column determined in the three samples as a function of stellar mass for for R$<$ 150 kpc.
The lower column density sample of Stocke-Bowen is about $10^2$ lower than the 
median of the COS-Halos sample where the samples overlap in stellar mass.  
The higher column density population may be similar between the two samples.
The numbers next to the highest HI column COS-Halos points show the impact parameter in kpc, 
where we note that most points lie within 50 kpc of the galaxy except for three systems at 83--91 kpc.
The highest column point in the Stocke-Bowen sample has an impact parameter of 53 kpc.
}
\label{fig:NHIR150kpcSBW}
\end{figure}

The distribution of N(HI) can have large uncertainties in the Stocke-Bowen sample, 
but when examining N(HI) as a function of impact parameter (Figure ~\ref{fig:NHIRSB}),
there is a bimodal distribution of the best-fit values, with a high column density 
grouping near $10^{18}$ cm$^{-2}$ and a lower column near $10^{13.5} - 10^{14.5}$ cm$^{-2}$.  
There is no dependence of this bimodal distribution on stellar mass (Figure ~\ref{fig:NHIR150kpcSBW}). 
All of the high column density systems have an impact parameter of 53 kpc or less and 
5/7 absorption line systems within this impact parameter belong to the high column density family. 
In contrast, 0/9 high column systems lie beyond 53 kpc. 

This bimodal distribution in columns can be interpreted as a disk and a halo phenomenon.  
Warm and cool gas is found in disks in many galaxies and the disk extends beyond 
the optical galaxy (e.g., \citealt{sancisi08}). 
Generally, the HI disk properties can be studied from 21 cm emission for columns in 
excess of $10^{19.5}$ cm$^{-2}$, but there are a few studies that probe more deeply.  
\citet{Pisano2014}, using the Green Bank Telescope (GBT), studied NGC 2997 and 
NGC 6946, both of similar mass to the Milky Way and M31, to a limit of N(HI) $\sim 10^{18}$ cm$^{-2}$.  
For NGC 2997, the mean outer bound for HI detection is 50 kpc, while for NGC 6946, 
the outer HI boundary is about 45 kpc, but there is a filament to the NW that extends to about 80 kpc. 
Another L* galaxy, NGC 2903, was observed with the Arecibo Observatory to a limiting 
HI column of  $2 \times 10^{17}$ cm$^{-2}$, where the HI disk has a radial extent of 
about 60 kpc along the major axis and 40 kpc along the minor axis \citep {Pisano2014}.  
Around more massive galaxies (logM$_{*} > 10^{11}$ M$_{\odot}$), an ongoing survey 
reports extended gas at distances of 50--100 kpc for about half of their nearby target 
galaxies, observed with the GBT \citep {Ford2016}.  
It appears that HI disk commonly extends to about 50 kpc around L* or more luminous 
galaxies for N(HI) detection sensitivities of $\sim 10^{18}$ cm$^{-2}$ or lower.

The HI columns found by \citet{stocke13} are below this value and they are also larger in radius than typical higher column density 21 cm disks \citep{roberts94,sancisi08}.  This suggests that lower N(HI) gaseous disks can extend to about 50 kpc in this sample of galaxies, some of which are small and of low luminosity. The absorption systems with column densities less than 10$^{16}$ cm$^{-2}$ could be the counterparts of halo clouds seen in the Milky Way, which have a net mass significantly lower than the disk gaseous mass. 

When examining N(HI) in the COS-Halos sample, the data may be consistent with there 
being two groups, one centered near logN(HI) = 16.3 and a second group of seven objects 
that with $18.2 < log(NHI) < 19.8$ (Figure ~\ref{fig:NHIR150kpcSBW}).  
One can test whether a distribution is consistent with bimodality \citep{knapp2007}, 
and using the method of Pearson, the sample formally meets the criteria for being bimodal, 
although we do not find this to be compelling. 

There are significant differences in N(HI) between the COS-Halos and Stocke-Bowen samples, 
much of which can be traced to the differences in equivalent widths.  
The lower column density group from COS-Halos (N(HI) $< 10^{18}$ cm$^{-2}$) has a median 
significantly higher than the lower column density group in the Stocke-Bowen samples 
(N(HI) $<10^{16}$ cm$^{-2}$) by about two orders of magnitude (Figure ~\ref{fig:NHIR150kpcSBW}).  
This difference persists in the luminosity region where the two samples overlap.
Most of this $10^2$ difference is due to systematically higher equivalent widths 
(a factor of $10^{1.7}$ in the column), with the remainder, a factor of two, due 
either to galaxy evolution or differences in the methods used to convert equivalent widths to column densities.
For the highest column density groups, the median value in the Stocke-Bowen sample is 17.9, 
while for the COS-Halos sample, it is 19.2, a difference of about a factor of 20.  
This difference is primarily due to the large equivalent width values of galaxies with 
optical luminosities above those of the Stocke-Bowen sample (Figure ~\ref{fig:NHIR150kpcSBW}).

Within the COS-Halos sample, the higher column density group is distinguished by the 
stellar mass of the galaxy.  The eight highest HI column systems have a median stellar mass 
of $1 \times 10^{11}$ M$_{\odot}$ whereas the hosts for the lower column systems have a 
stellar mass that is three times lower, $3 \times 10^{10}$ M$_{\odot}$.

\begin{figure}       %%%%%%%%%%%%% Figure 17 (was 17)
\centering
\plotone {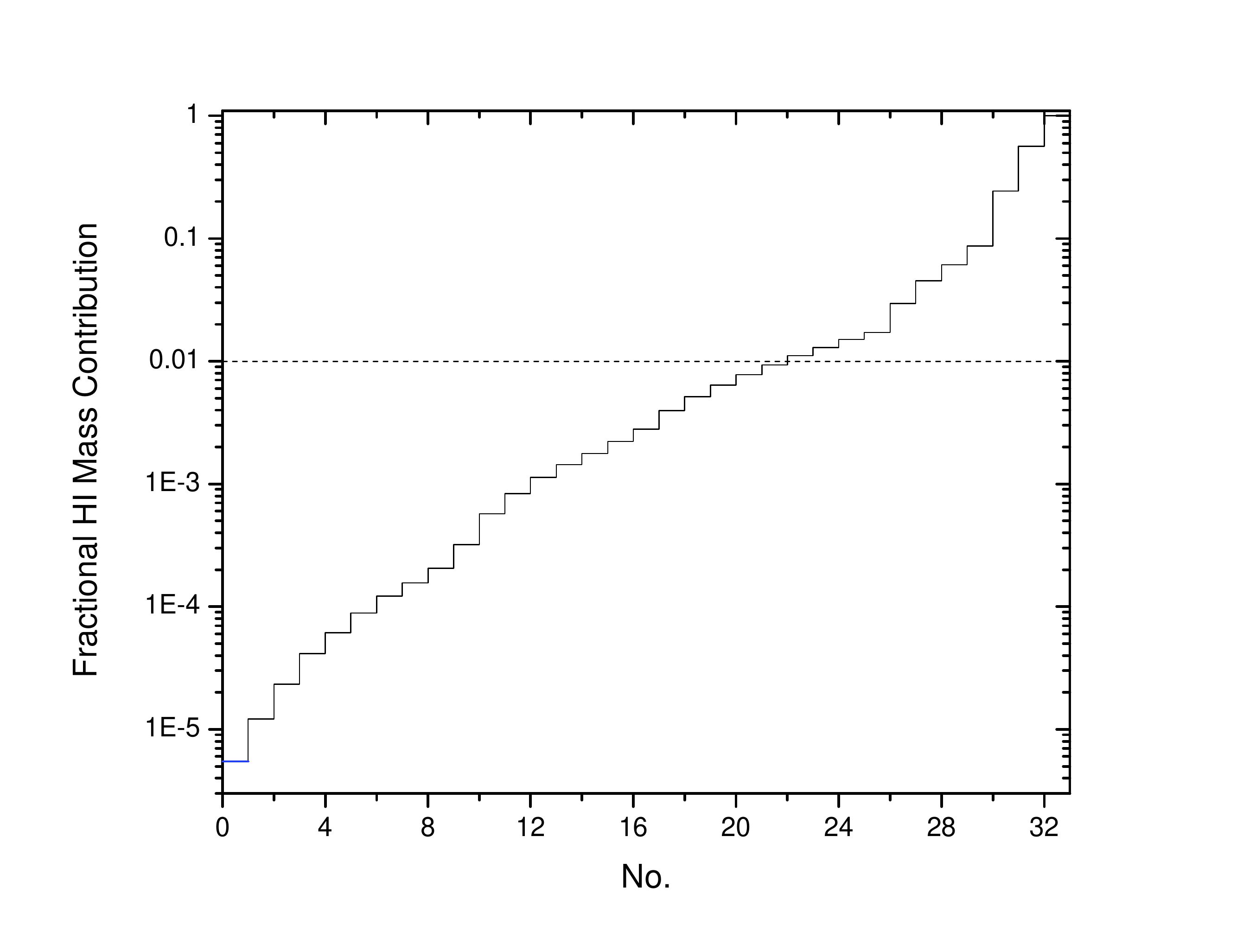} 
\caption{
The cummulative sum of N(HI)R$^2$ from the COS-Halos sample, normalized to unity. 
The systems with the two largest values of N(HI)R$^2$ account for 75\% of the the total amount and 
the top six systems account for 97\% of the total amount. 
}
\label{fig:FractionalHIMasses}
\end{figure}

When considering the radial distribution of the high column density group in the COS-Halos 
sample, 5/8 lie within 50 kpc of the target galaxy (Figure ~\ref{fig:NHIR150kpcSBW}), 
which can be understood as absorption by an extended gaseous disk.  
However, three absorption systems lie at about twice the distance of the inner 
group, 83--91 kpc from the target galaxy. 
We can obtain a rough estimate of the HI mass contributions from the various systems 
by calculating the product of N(HI) and the square of the impact parameter.  
In doing this, we see that just a few systems account for most of the HI mass.  
Two systems with impact parameters of about 83 kpc and 91 kpc (logN(HI) of 19.6 and 19.5) 
account for about 75\% of the HI mass and just six systems account for 97\% of the HI mass (Figure ~\ref{fig:FractionalHIMasses}).  
We inspect these systems individually to understand if there is something special about their nature.

\begin{figure}       %%%%%%%%%%%%% Figure 18 (was 18)
\centering
\epsscale{0.8}
\plotone {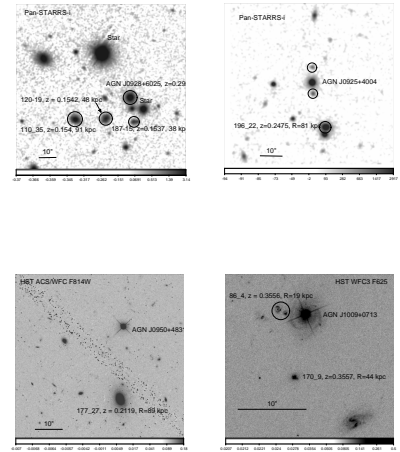} 
\caption{
Optical images of four of the AGN-galaxy pairs that account for most of the HI mass, 
with the top two being the most important.
The absorption attributed to 10\_35 (top left) may be due to blue spirals closer to the AGN.
Around the AGN J0925+4004 (top right), there are two nearby objects (circles, no redshifts) 
that are potential absorbers.
Near AGN J0950+483 (bottom left) there are no obvious galaxies much closer than 177\_27, 
except for the galaxy to the east with a separation of 72 kpc, if at the same redshift.
Absorption against AGN J1009+0713 (bottom right) is assigned to galaxy 170\_9, but it could be 
due to the two small galaxies projected 15--20 kpc from the AGN.
}
\label{fig:fourfields}
\end{figure}

The two largest potential contributors to the net HI mass (the largest values 
of N(HI)R$^2$) are the galaxy identifications 196\_22 and 110\_35, which we examine 
in more detail using archival images, including redshifts from SDSS (Figure ~\ref{fig:fourfields}). 
The system 196\_22 ($z_{gal}$ = 0.2475, R = 83 kpc; AGN J0925+4004) has two small objects close to the AGN.  
This raises the concern that these closer lower luminosity objects could be the absorbing galaxies.  
Spectroscopic observations are needed to address this issue.  
The 110\_35 system ($z_{gal}$ = 0.154, R = 91 kpc; AGN J0928+6025 at z = 0.29) is part of 
a group of galaxies, of which two members are closer to the AGN (38 kpc and 48 kpc impact parameter). 
These two members are active star forming systems, being relatively luminous in the UV \textit{Galex} 
images, whereas the galaxy assigned the absorption (110\_35) is barely visible in the UV.  
It seems possible, if not likely, that the true absorbing systems are the spiral galaxies closer to the AGN.  
There is also a group of galaxies at the redshift of the AGN, which makes the image complex. 

The third high column system with a large separation is 177\_27 (z = 0.2119, R = 91 kpc; AGN J0950+483), which does not have any moderately bright galaxies closer than 50 kpc. There is another galaxy to the east, which would have an impact parameter of 72 kpc if it were at the same redshift. Another system, galaxy 170\_9 (z = 0.3557; sixth largest value of N(HI)R$^2$) is only 44 kpc away from the AGN J1009+713, suggesting a true association.  However, a high-resolution image shows two galaxies at a separation of 15 kpc and 20 kpc and with the same redshift as 170\_9.  It might not be possible to determine which is the true absorber without complementary HI maps, which are far beyond current capabilities. 

This examination of the high N(HI) systems at impact parameters $> 50$ kpc suggests that two of the three systems (196\_22 and 110\_35) may have absorption attributable to nearer less luminous galaxies.  Excluding these two systems would decrease the net total HI mass by a factor of four.  
HI is occasionally seen distributed through galaxy groups at $z \approx 0$, such as in the Leo Group, so if this occurs at $z \approx 0.2$, it could explain the occasional high column absorption far from a galaxy. 
 
The HI masses can be inferred by adopting a characteristic column and radius, where the mass is given by 
M(HI) $= 6.2 \times 10^7 (N(HI)/10^{18}) (R/50$ kpc)$^2 $ M$_{\odot}$.
Of the high column density group of Stocke, a median system has a mass of $3.5 \times 10^{7}$ M$_{\odot}$ out to 50 kpc. Only about 20\% of the galaxies have high columns, so when averaged over all galaxies, the value would decrease to $7 \times 10^{6}$ M$_{\odot}$. 
For the lower column density absorbers in the Stocke-Bowen samples, the peak of the distribution occurs at log N(HI) = 14.3, so M(HI) = $1 \times 10^{5}$ M$_{\odot}$ out to 150 kpc.  
In the COS-Halos survey, the high column density systems have a median near logN(HI) = 19.3, leading to a HI mass of $1.2 \times 10^{9}$ M$_{\odot}$ out to 50 kpc and $5 \times 10^{9}$ M$_{\odot}$ out to 100 kpc.  As these systems comprise about 20\% of the sample, the mean mass would be five times lower.
The lower column density systems have a median N(HI) about two orders of magnitude larger than that in the Stocke-Bowen study, so the masses of this component are  $1 \times 10^{7}$ M$_{\odot}$ out to 150 kpc. 

\subsection{Total Hydrogen Columns}

A calculation of the total absorbing mass requires a very significant ionization correction. This correction is nearly always obtained by adopting a photoionization model for the gas clouds.  The properties of such absorbing clouds are poorly known as they are not detected in emission, so sizes, densities, and filling factors are not independently known.  However, the properties of the incident radiation field are estimated from the ensemble of AGNs and leakage from UV-bright galaxies (e.g., \citealt{haardt2012}), so a successful photoionization model fit yields the gas density and temperature, from which one can determine the gas pressure and the cloud size.  The resulting cloud properties should be consistent with the ambient pressure expected in the halo (pressure equilibrium) and the observed sizes of clouds around the Milky Way, for example.

From the analysis for the COS-Halos sample \citep{werk2014}, the pressure (expressed as nT) is surprisingly low both for the high and low column samples, with median values of 4 K cm$^{-3}$ and 1 K cm$^{-3}$ , respectively.  The characteristic pressure of a virialized system is $n_{200} T_{vir} \approx 70 (T/3 \times 10^6 K)$ K cm$^{-3}$, much larger than the values determined from the photoionization model.  The pressures inferred for the Milky Way hot halo are somewhat larger within 150 kpc \citep{salem15, faerman16}, generally more than two orders of magnitude larger than those from the photoionization analysis (for an incident photon flux of 3000 photons cm$^{2}$ s$^{-1}$).  If one were to demand that the gas clouds have a pressure characteristic of their location in the halo, the flux of ionizing photons would be too small by the same large amount.

Another concern, also first identified by the COS-Halos team \citep{werk2014}, is that the clouds and their masses can be enormous.  Half of the clouds (16/33) have inferred lengths greater than 100 kpc, with two greater than 1 Mpc.  About one-quarter have masses comparable to or larger than the stellar masses, in some cases by large amounts.  Several in this group would have a baryon mass exceeding the cosmological value of the host system.  These problems raise concerns with the photoionization analysis, which may reduce the reliability of the ionization correction that one would apply to the HI column to obtain the total hydrogen column.

Given these concerns, we adopt a median value for the conversion of N(HI) to N(H), based on the photoionization fits by \citep{werk2014}.
This leads to N(H)/N(HI) $\approx 400$ for the lower N(HI) systems. 
The resulting values of N(H) are similar to what would be obtained by multiplying the metal column for an ion (e.g., C, Si) by the metallicity correction, for a metallicity of about 0.2 solar.  
This conversion factor raises the gaseous mass for this ensemble of clouds to about $5 \times 10^{9}$ M$_{\odot}$ out to 150 kpc and a median column per cloud of $6 \times 10^{18}$ cm$^{-2}$.  
For the higher N(HI) systems, the median conversion factor is about 6, so the hydrogen masses of these systems are $7 \times 10^{9}$ M$_{\odot}$ out to 50 kpc and $3 \times 10^{10}$ M$_{\odot}$ out to 100 kpc (median column of $1 \times 10^{20}$ cm$^{-2}$).  If averaged over the total number of galaxies, these masses are lowered by five, as only 20\% of the galaxies have high columns.  However, these high N(HI) systems occur in the highest mass galaxies, and within that group, about half of the galaxies show such systems.  Therefore for a typical high-mass galaxy (10.8 $<$ log$M_*$ $<$ 11.5), these masses should only be lowered by about a factor of two.

For the targeted galaxy sample of  \citet{stocke13}, they calculate photionization corrections for a subset and obtain N(HI) to N(H) conversion factors that are very similar to \citet{werk2014}, with a median value of 450 and a lower conversion for two higher column density systems.  For consistency, we use the same conversion factors as above and find that the low HI column sample clouds have a mass of $5 \times 10^{7}$ M$_{\odot}$ out to 150 kpc, while the higher N(HI) group have hydrogen masses of $2 \times 10^{8}$ M$_{\odot}$ out to 50 kpc. 

To summarize, the absorption line systems around targeted galaxies is consistent with being bimodal in N(HI) and can be understood as caused by two components: a halo of clouds that extends to at least 150 kpc; and an extended higher column disk of gas that extends to 50 kpc and in rare cases, to 100 kpc. We propose metallicity and ionization conversion factors which, if correct, point to a picture where the mass of halo clouds is $\sim 5 \times 10^{9}$ M$_{\odot}$ (COS-Halos sample), while the extended disk has a mass of about $\sim 3 \times 10^{9}$ M$_{\odot}$ out to 50 kpc ($\sim 1 \times 10^{10}$ M$_{\odot}$ out to 100 kpc).  These extended halos are most frequently found in galaxies with $M_*$ $\gtrsim 10^{10.8}$ M$_{\odot}$.  While these halo and disk gas masses are considerable, they are at least an order of magnitude less than the missing baryons, typically $\sim 10^{11}$ M$_{\odot}$ for an L* galaxy. Next we will explore further implications of our assumed metallicity and ionization conversion factors, and show that they satisfy other physical and observational constraints, such as producing the correct number of absorbing clouds and satisfying pressure equilibrium with the hot halo.

\subsection{Number of Clouds in the Halo}

The number of clouds and their sizes can be estimated through various approaches 
(e.g., \citealt{stocke13}) and here we offer another method.  We suppose that the 
absorbing clouds are in pressure equilibrium with the hot ambient medium, which leads 
to a density of $n_{cloud} \sim n_{200} T_{vir}$/$T_{cloud}$.  
For a Milky Way type galaxy ($T_{vir} \approx 10^{6.3}$ K) and the usual temperature of 
a photoionized cloud (10$^4$ K), n$_{cloud} \sim 10^{-2}$ cm$^{-3}$.  
We calculated a characteristic halo cloud hydrogen column in the COS-Halos sample of 
$6 \times 10^{18}$ cm$^{-2}$ (above), and for a spherical uniform cloud of radius $r_c$, 
the characteristic path length is $2^{1/2} r_c$, leading to $r_c \approx 5 \times 10^{20}$ cm 
(160 pc), and a cloud mass of 10$^{3.6}$ M$_{\odot}$.  This implies that a typical L* galaxy halo 
contains $\sim 10^6$ clouds.  If such a cloud were in the Milky Way halo, at a distance 
of 10 kpc (the HVC), the diameter would be $\sim 2^{\circ}$, similar in size to some HVC 
and the substructures seen in large HVC complexes.  At these higher densities, photoionization
would appear to be ineffective, but there is a successful alternative ionization model in which
the ionization is driven by the turbulence and ultimately the motion of the clouds \citep{gray15}.

This calculation assumed that a single cloud produces the absorption, but multiple components are common around a single galaxy \citep{stocke13,werk2014}.  We can estimate the number of components along a galaxy site line by using the covering factor or counting components and we arrive at about the same result.  For equal size clouds and a covering factor of 90\% \citep{werk2014}, there would need to be an average of 2.4 clouds per line of sight in order for zero clouds to be observed 10\% of the time (using Poisson statistics).  Alternatively, one can try to count the number of components observed, although the ability to identify components depends on the ion, the S/N,  and the optical depth of the feature. When we count individual components in the line profiles of \citet{werk13a}, we find an average of 2.2, and this would appear similar to that found by \citet{stocke13} in their targeted survey.  For 2.3 clouds along the line of sight, and in pressure equilibrium as given above, the cloud radius is $\approx 4 \times 10^{20}$ cm (130 pc), and with a cloud mass of $\sim 10^{3.3}$ M$_{\odot}$,
similar to the values from the previous argument. 
% like a "rain" of clouds, which may be formed and destroyed on a timescale shorter than the infall timescale
% double check the mass of the COS-Halos clouds

\section{Constraints on the Baryon Content from the SZ Measurements}

The thermal Sunyaev-Zeldovich effect is detected toward many rich galaxy
clusters, yielding valuable information on the gas properties of these
systems. \ As sensitivities improve, it opens the possibility that this method
can provide information about the hot gas properties of galaxy groups and luminous
galaxies, which we examine here.

The traditional Compton $y$ parameter along a line of sight is $y=(\sigma
_{T}/m_{e}c^{2})\int Pdl$, where $\sigma_{T}$ is the Thompson cross section,
$m_{e}$ is the electron mass, $c$ is the speed of light, and $\int Pdl$ is the
pressure integral along the line of sight. \ 
Clusters or galaxies are taken to be extended and treated as spherical objects
at some distance $D$, so an integrated signal within a single beam is given by
$Y_{500}=(\sigma_{T}/m_{e}c^{2})kD^{-2}\int_{0}^{R_{500}}4\pi r^{2}n_{e}Tdr$,
where $k$ is the Boltzmann constant and the subscript 500 denotes the value at
the radius for which the overdensity of matter reaches 500 (e.g., \citealt{arnaud10}). 
For the typical
convention where the virial radius is given by $R_{200}$, $R_{500}%
\approx0.7R_{200}$. 
For clusters of galaxies, where the density steepens and
the temperature decreases significantly before the virial radius, $Y_{500}$ is
only 2\% lower than the value that would be obtained if integrating to
infinity, so this is a useful observational quantity. \ However, for galaxies,
the temperature structure is not known near $R_{500}$ so it could be falling less
rapidly. \ Also, where the density can be measured, it has a shallower
radial decrease than in clusters, so the contribution to the $Y$ parameter at
larger radii is likely to be greater \citep{lebrun15}. \ In the case where the temperature is
constant within the region of significance (to $Y$), the above parameter
becomes $Y_{500}=(k\sigma_{T}/m_{e}^{2}c^{2})TM_{e}(R_{500})D^{-2}$, where
$M_{e}(R_{500})$ is the mass of electrons within $R_{500}$, but as we shall
see, it will be necessary to consider larger radii for the case of galaxies.
\ The value of $Y_{500}$ is typically given in square arcminutes.

\subsection{Planck SZ Detections Of Stacked Galaxies}

The most favorable galaxies for detection are massive ones, and since they are
uncommon, the \citet{planckXI} stacked many galaxies and found a signal using the
100--353 GHz bands, which has an angular resolution of about 10$^{\prime}$.
\ They bin their data by the mass of the stellar content, M$_{\ast}$, which
they obtain from the catalog of \citet{blanton05} that is based on the
SDSS survey galaxies. \ Approximately 260,000 galaxies are used in the SZ study,
although above logM$_*$ = 11.0, there are 58,000 galaxies.\ They clearly 
detect the SZ effect for galaxies of mass logM$_{\ast} \geq 11.25$ and it 
is likely that they detect the effect somewhat below that value. 

An independent effort, using a similar approach, was taken by Greco et al.
(2014). \ Their screening is a bit different and they consider an additional correction, 
from the emission of dust grains within the galaxies, which makes 
only a modest difference. \ They smooth all frequency bands to 10$^{\prime}$ 
resolution and then use aperture photometry, extracting a signal within 5R$_{500}$. 
This procedure differs from that of \citet{planckXI}, who left each map at the 
native resolution and fit functional forms of the pressure distribution.  
The resulting amplitude of $Y_{500}$ is similar to that of 
\citet{planckXI}, although their S/N is lower, so their signal is significant 
beginning at a slightly higher mass range.

\subsection{Expected SZ Signal From X-Ray Observations of Massive Galaxies} 

\begin{figure}       %%%%%%%%%%%%% Figure 
\centering
%\epsscale{0.8}
\plotone {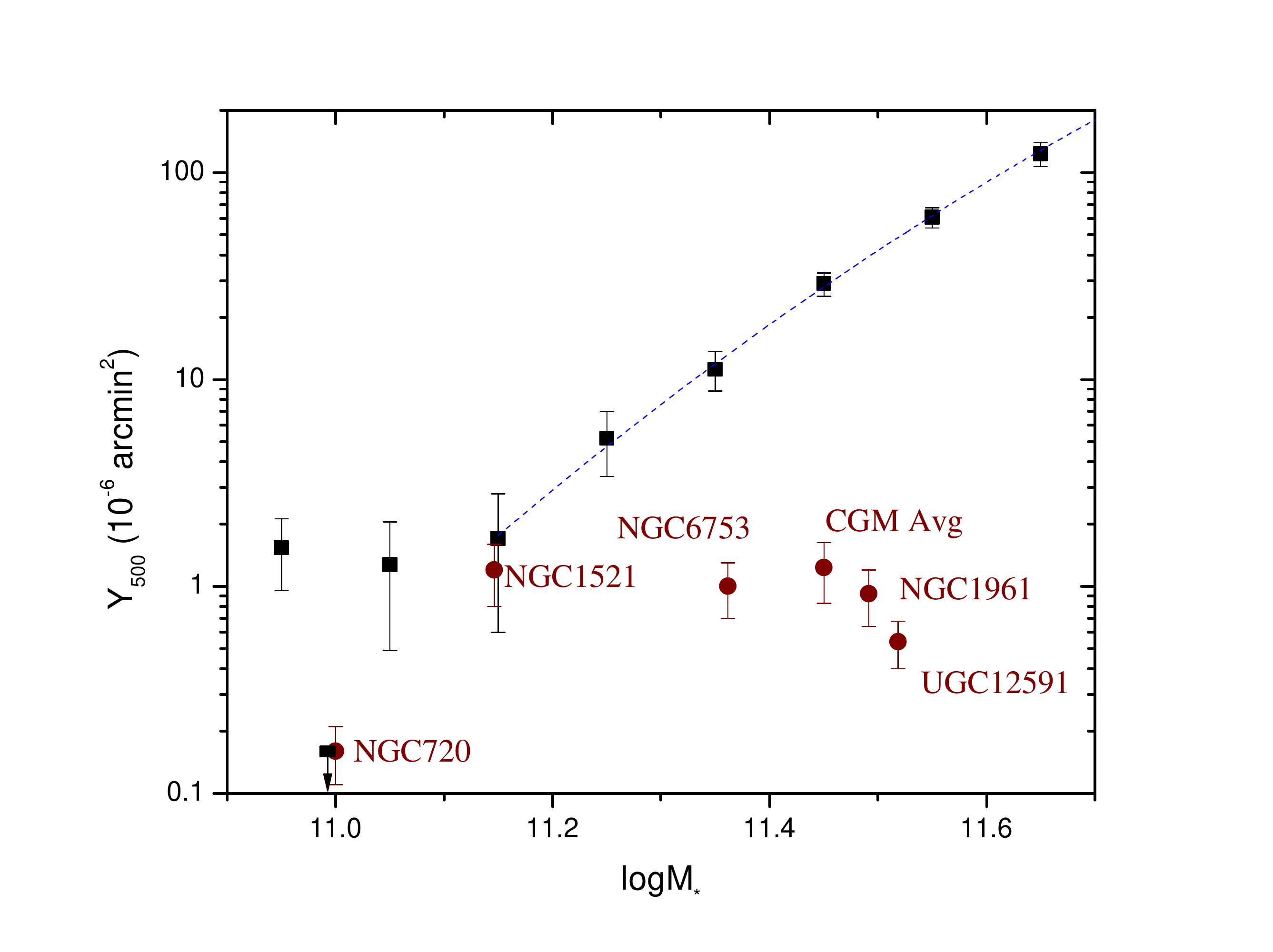}
\caption{
The SZ Y parameter, normalized to a distance of 500 Mpc for the stacked galaxies \citep{planckXI} and for the Y parameter inferred from the X-ray properties of five nearby massive galaxies (Table 1), plus a model galaxy that represents the average of the CGM-MASS sample \citep{jiangtao17,jiangtao18}.  For the two lower mass galaxies, both of early-type (NGC 720 and NGC 1521), the $Y_{500}$ values are consistent with an extrapolation of the relationship from  \citet{planckXI}. At higher mass, the galaxies (all spirals) lie at least an order of magnitude below that relationship.   For the spiral galaxies, the gas masses were calculated assuming 0.25 Solar metallicity, and the error bars reflect a factor of two uncertainty in metallicity plus the uncertainties in extrapolating the gas mass to R$_{200}$. 
}
\label{fig:SZcomparison}
\end{figure}

There are few individual massive isolated galaxies for which there is good
X-ray data of the hot gas, with the best cases listed in Table 1; supplementing this is
the ongoing CGM-MASS sample \citep{jiangtao16,jiangtao17,jiangtao18}.
We can use these observation to calculate the SZ Y parameter for comparison with the stacked results from Planck \citep{planckXI}, which gives Y as a function of stellar mass (Figure ~\ref{fig:SZcomparison}).
When calculating the SZ Y parameter for massive isolated galaxies with extended hot halos, we used the most recent gas mass and temperature determination (Table 1), although our result is insensitive to the choice of investigator results.  The gas mass depends on the square root of the metallicity (approximately), but as the metallicity for the spiral galaxies has large errors, we use a value of 0.25 Solar, which is within a factor of two of the range of measured values.  For the early type galaxies, NGC 720 and NGC 1521, we adopted the measured values, typically 0.4--0.6 Solar (Table 1).  Throughout an individual halo, we assumed a constant metallicity, temperature, and density power-law index.  If the temperature decreases with radius, as is seen in galaxy clusters, our values of  $Y_{500}$ would be larger than their true values.  A lower metallicity raises $Y_{500}$, with a factor of two decrease in metallicity raising $Y_{500}$ by 40\%.  Neither of these uncertainties modify the result (Figure ~\ref{fig:SZcomparison}) that the higher mass galaxies have values of $Y_{500}$ at least an order of magnitude below the stacked results \citep{planckXI}.  The value of $Y_{500}$ increases if the missing baryons are hot (either within or beyond R$_{200}$), with a typical increase in $Y_{500}$ of a factor of three.  This still is insufficient to account for the difference relative to the Planck results.

We consider the reasons for this discrepancy.
One obtains the predicted value of $Y$ from the observed relationships between the 
stellar luminosity and $Y$ in large stacked data sets.  For the stellar mass range of interest, about $10^{11.5} M_{\odot}$, both the stellar luminosity and $Y$ are well-measured, with uncertainties of less than 20\%. 
The limit to the expected signal from individual galaxies is based on the maximum possible mass of gas, which in turn is inferred from the halo mass, the cosmological value of the baryon fraction, and the amount of baryons in stars and cold gas. 
The halo mass is inferred from the flat part of the rotation curve, coupled to a NFW profile for the dark matter distribution.  For the halo mass to be incorrect by an order of magnitude, the rotation curve of NGC 1961 would exceed 800 km s$^{-1}$, which is never observed in the outer parts of galaxies.
Another possibility is that the temperature rises beyond the radius at which the X-ray observations can no longer determine the temperature.  However, if the temperature rose by even a factor of two, the gas would be unbound from the galaxy and flow outward, cooling by adiabatic expansion and decreasing the value of $Y$.
It is likely that the gas temperature decreases with radius, in which case we have overestimated the maximum value of $Y$ through our assumption of isothermality.

There have been some issues raised regarding the values of $Y_{500}$ published by the \citet{planckXI}.  They state that the $Y_{500}$ - $M_h$ relationship is self-similar, implying that the hot gas properties around a cluster of galaxies is similar to an individual galaxy.
In particular, self-similarity implies the same shape for the density and temperature, as well as the same fractional baryon mass as a hot component.  This is questioned by \citet{greco15} and \citet{lebrun15}, who point out that the observed density profiles are flatter in individual galaxies than in galaxy clusters.  Also, feedback will have a larger effect for galaxies than rich clusters.  These arguments suggest that one might use a pressure profile for galaxies that is flatter than for clusters when converting the $Y$ signal within $5R_{500}$ to the value within $R_{500}$.  While these are all sensible considerations, it leads to modest differences in $Y_{500}$ for the galaxies of interest \citep{greco15}.

There may be other issues introduced by the observational requirement that the stellar mass is an independent variable. \citet{wang2016} used new weak lensing observations to estimate the effective halo masses of each stellar mass bin in the \citet{planckXI} sample, and showed that there is both significant dispersion and significant model-dependence in the distribution of halo masses as a function of stellar mass. With the weak lensing data, they were able to account for some of this model dependence, and renormalize the effective halo masses of each bin. This renormalization brought the $L-M$ and $Y-M$ relations in the stacking sample into agreement with the relations observed for galaxy clusters.

However, while \citet{wang2016} showed that there is very little uncertainty in the behavior of mean properties in the stellar-to-halo mass relation (see also e.g. \citealt{moster10,behr2010}),  this is not guaranteed for the behavior of outliers. The fraction of elliptical and lenticular galaxies rises extremely sharply at stellar masses above $10^{11}$ M$_{\odot}$ (e.g. \citealt{bern2010}) and even among spiral galaxies, nearly all of them at the stellar mass of NGC 1961 are passive \citep{wilman2012}, so a massive and moderately star-forming spiral galaxy like NGC 1961 is extremely unusual.  Indeed, NGC 1961 is one of the most massive such galaxies known in the local Universe. 

Weak lensing studies such as \citet{velander14} show that for spirals, there is a nearly linear relationship between $M_h$ and $M_{*}$, while for the red galaxies, $M_h \propto M_{*}^{1.36}$, so the halo mass grows more rapidly with the stellar mass. For a stellar mass of $10^{11.5} M_{\odot}$ the red galaxies have a halo mass twice that of the blue galaxies. For a $Y-M$ relation with a slope of 1.61 \citep{wang2016}, this translates to a factor of 3 in $Y$, which reduces the tension with the Planck data, but does not wholly resolve the issue. Again, however, this observation relies on mean properties. If NGC 1961 is a factor of two below the mean halo mass relation for spiral galaxies then the implied reduction in $Y$ is a factor of 9.3 compared to the mean for ellipticals, which would largely resolve the discrepancy.

Fundamentally, the issue is probably how to compare results about isolated galaxies to results about central galaxies, even when the galaxies have the same stellar mass. Most of the X-ray results for hot halos are derived from studies of isolated galaxies, while the SZ results are measured for stacks of central galaxies. Numerical simulations can help to connect these two types of selection criteria, but it would be extremely informative to have observations that bridge this divide. This can include studies of larger samples of outliers, such as the CGM-MASS sample \citep{jiangtao16,jiangtao17}, with moderate X-ray observations of six additional isolated giant spirals. Stacking of the SZ signal from samples of isolated galaxies, or deep SZ observations of individual systems, is also necessary to complement the X-ray results.

Another consideration is whether there is observational evidence for a significant amount of hot gas beyond the virial radius, either due to a group medium or to accretion filaments.  
The angular extent of the SZ signal suggests that galaxy groups or poor clusters 
may contribute.  A luminous galaxy like NGC 1961 ($M_* = 3.1\times10^{11}M_{\odot}$)
has a virial radius of about 470 kpc and at the mean distance of the sample, 500 Mpc,
this subtends a diameter of 6.0$^{\prime}$, which is less than the FWHM of the instrument, 10$^{\prime}$.
Therefore, the galaxies should appear as point sources, but for the stacked images
in the bins centered at log$M_*$ = 11.15, 11.25, the emission is somewhat extended \citep{planckXI,greco15}.
The extent of the SZ signal is studied further by \citet{vanW14} and \citet{ma2014}, who use their weak galaxy lensing survey and perform a cross-correlation with the Planck data.  Although their signal is at the 4$\sigma$ level, they find that about one-third of the SZ signal comes from beyond the virial radius in their $M_h = 10^{12} - 10^{14} M_{\odot}$ bin and one half of the signal is beyond the virial radius for the higher mass bin, $M_h = 10^{14} - 10^{16} M_{\odot}$. 
This extended nature of the ionized gas is supported by kinetic SZ studies \citep{planckXXXVII,hern2015}, although the signal is weaker than the thermal SZ investigations, so the constraints are poorer.

To conclude, the SZ signal seen toward a set of stacked locally brightest galaxies 
suggests that a significant fraction of the galactic baryons are hot.  
However, for the observed stellar mass of NGC 1961 and other similar nearby 
massive spiral galaxies (and one early-type galaxy), the expected SZ signal 
is at least an order of magnitude below that inferred from the stacked ensemble.  
A resolution to this discrepancy may have a few components, such as the difference 
in the halo mass of early and late type galaxies of the same stellar mass, and
a SZ contribution from hot gas beyond $R_{200}$ but gravitationally associated with the 
galaxy (within the turnaround radius).
However, it is possible that the selected high-mass spirals are not typical of the 
stacked galaxy sample from which the SZ signal is extracted. 
This suggestion can be examined further, such as extracting SZ signals from 
galaxy samples sorted by morphology or color.

\section{Discussion and Summary}

Several investigators have used different observations and conclude that 
they have found the “missing” baryons in individual galaxies, but some 
of these claims are mutually incompatible. The COS-Halos team use their 
UV absorption line observations of warm ionized gas and argue that this 
component, within $R_{200}$, completes the baryon census for galaxies. 
For the phase with T $\sim$ T$_{virial}$, \citet{faerman16} offers a 
model for the Milky Way where the hot and warm baryons within $R_{200}$ completes the baryon census. 
\citet{gupta12} has modeled X-ray absorption and emission line observations, 
with extrapolations, to account for the all Milky Way baryons within $R_{200}$. 
\citet{nica2016} combines O VII absorption in the disk and halo, 
arriving at a yet different model for the hot gas distribution in the Milky Way, and with 
a hot gas mass that accounts for the missing baryons within about $R_{200}$.
We argued that the hot gas mass within $R_{200}$ does not account for the 
missing baryons in the Milky Way ($\S$2.5). 
We find a similar result for external galaxies and conclude that to account 
for the missing baryons  ($\S$2.6), the hot halo would have to extend beyond $R_{200}$.  
Finally, SZ studies with Planck detect signals that extend beyond 
$R_{200}$ and the signal is easily strong enough to account for the 
missing baryons as hot gas, if the metallicity is low enough that 
X-ray luminosity limits are not violated ($\S$4).  
Several of these results are in conflict with each other.  
Sorting out this situation was a primary motivation for this work.

Hot halos, as studied through X-ray emitting and absorbing lines, 
are fairly well understood for R $\lesssim 50$ kpc, where the temperature 
is about 50-100\% hotter than the virial temperature, $n \propto r^{-3/2}$ ($\beta = 1/2$), 
and the metallicity is about 0.1--0.5 solar, except for early-type 
galaxies where the metallicity is about solar.  
The mass within 50 kpc is $\sim 10^{9.5}$ M$_{\odot}$, so determining the
mass out to $R_{200}$ (250 kpc) requires an extrapolation with a density model.  
Flattened density models ($\beta \lesssim 0.3$) lead to large gas masses, which 
if correct, could contain the missing baryons within R$_{200}$. 
One such model was suggested by \citet{nica2016}, but we showed that this is
in conflict with observed emission line data, along with other issues ($\S$2.3).
The model of \citet{faerman16} cannot be ruled out, as the density law is consistent
with most existing Milky Way observations ($\S$2.5). 

One can consider whether the density distribution of \citet{faerman16} 
is expected from models of galaxy assembly.  
Simulations of diffuse coronae in galaxies rarely show a flattening to the 
hot gas component with radius (e.g., \citealt{soko2016}) nor is it 
expected from general formation of structure considerations \citep{tozzi2001}.  
Structure formation models do not suggest the density distribution of \citet{faerman16}.

%In estimating the mass of the hot halo, we used a $\beta = 1/2$ density law, 
%which is consistent with existing observations for $r \lesssim 50$ kpc.  
%An extrapolation of the $\beta = 1/2$ density law to $R_{200}$ leads to
%a hot gas mass that is comparable to stellar mass, but about half of the baryons are still missing.

Single galaxies do not produce a detectable SZ signal ($\S$4), so stacks of 
galaxies were used, which produces a signal for massive galaxies (log$M_* > 11.2$).  
These results indicate that much, if not all of the missing baryons are hot and 
are extended beyond R$_{200}$.
However, the observed signal is too large for the amount of baryons expected 
in the halos of giant spiral galaxies, inferred from X-ray observations. 
A possible resolution of this discrepancy is that there could be substantial 
differences in the hot gas halos around massive spirals compared to massive early type galaxies. 
Until such issues are resolved, it will be difficult to use the SZ results to 
constrain the hot gas content of spiral galaxies.

\subsection{UV Absorption Line Studies and a Disk-Halo Model}

The UV absorption line studies are extensive and we have added to the 
analysis ($\S$3), showing that there is a decrease in the equivalent width 
and gas column density between $z \approx 0.2$ and $z \approx 0$ 
(between the COS-Halos and Stocke-Bowen samples).  
The larger HI equivalent widths between COS-Halos and the Stocke-Bowen sample 
is also present with the COS-GASS survey (median z =0.039; \citealt{bor2015,bor2016}), although to a lesser degree. 
For the region of overlap between COS-Halos and COS-GASS, impact parameters 
of 60--150 kpc, 9/10 of the highest H I equivalent width systems occur in the COS-Halos sample.  
The median of the COS-GASS sample is about 25\% below that of 
the COS-Halos sample, for the 60--150 kpc groups.
The H I equivalent width distribution of COS-GASS appears to be intermediate 
between COS-Halos and the Stocke-Bowen sample, where there is a factor 
of two difference in the median equivalent width values.

One does expect a difference in equivalent widths, due to evolution of gas in 
galaxies with cosmic time.  The mass of dust at low redshift rises as a steep 
function of redshift for spiral galaxies in the \textit{Herschel}-ATLAS survey, 
M$_{dust} \propto (1+z)^p$, where $p \approx 4$ \citep{Bourne2012}.  
Assuming that the dust mass is linearly proportional to the dust-bearing gas mass 
(mainly gas with T $< 10^{5.5}$ K), we see that a difference between z = 0.2 and 
z = 0 corresponds to a decrease in the gas mass of a factor of two. 
This leads to a decrease in the equivalent width of about 25\% for lines on the 
flat part of the curve of growth.  
This decrease is consistent with the difference between the COS-Halos sample and the COS-GASS sample. 
However, the COS-GASS sample was selected for having a gaseous disk that was quantified 
with 21 cm observations, while the COS-Halos sample does not have this selection criteria.
Therefore, adjusting the equivalent widths between the two samples by the above method
may not be valid because of the difference in sample selection criteria. 
In a comparison between the COS-Halos and Stocke-Bowen samples, the equivalent width 
differences are significantly larger, the reason for which remains unclear.

We suggested that the absorption around galaxies can be interpreted as coming 
from two populations: a disk component, largely responsible for the 
high N(HI) columns within an impact radius of 50 kpc; and a more extended halo component.  
In the $z \approx 0.2$ study (COS-Halos), the mass of the disk component 
can be $5-10 \times 10^9$  M$_{\odot}$, but they only occur in about 20\% 
of the sightlines, whereas the halo clouds contribute about 
$5 \times 10^9$ M$_{\odot}$ in nearly all galaxies.  
This is significantly less than the estimate from the COS-Halos team but 
is similar to the value suggested in the analysis of \citet{keen17}. 

A model that has striking similarities to the H I absorption line data 
around galaxies is from the \textit{Illustris} simulation \citep{kauff2016}.  
In a model closest to the Milky Way (their Figure 14), they show a column 
distribution with radius that appears to be have two populations, one at 
about N(HI) = $10^{19}-10^{21}$ cm$^{-2}$ out to about 50 kpc, and another 
in a broad range,  N(HI) = $10^{13}-10^{17}$ cm$^{-2}$ that is steadily declining to R$_{200}$.  
Their model fails to reproduce the high columns near 100 kpc that are 
found in COS-Halos, but we have suggested that most of those high column 
detections at 80-92 kpc may not be associated with the target galaxies, 
which are much more massive than the Milky Way.  If one excludes those few 
large radius high column systems, the model and data appear consistent.  
The model could provide further insight if the absorption were identified 
with structures, such as a rotating disk of gas or halo clouds that may be 
falling in or flowing outward.

\subsection{Comparison of Disk and Halo Gas of Local Galaxies with Absorption Samples}

We compiled a comparison of the various extended gas mass determinations that appear in the
literature, including the Milky Way \citep{zheng15}, Andromeda \citep{lehner15,lehner17}, along with 
our analysis of the Stocke-Bowen and COS-Halos samples (Table 2; 
note that the disk refers to the extended disk and does not include the HI in the optical disk).
There are differences in the distances to which the gas is determined in each of these systems,
so we needed to correct or extrapolate the masses to a similar distance for final comparison.
To do so, we used the functional form of the density distribution for gas with T $< T_{virial}/4$
in the simulation of \citet{fielding16} for the galaxy with M$_{halo} = 10^{12}$ M$_{\odot}$
and where we took the average of the low and the high feedback cases.  
From this simulation 90\% of the halo gas lies within 0.33R$_{200}$ for the low feedback case and
0.45R$_{200}$ for the high feedback case. This would imply that if a significant fraction of
gas lies beyond these radii, it is most likely beyond R$_{200}$ and just projected onto the galaxy.
The correction from 0.2R$_{200}$ to R$_{200}$ is 2.3, while the correction fro 0.5R$_{200}$ to R$_{200}$
is just 5\%. 

This compilation shows that the Milky Way and Andromeda galaxies are similar and are similar
to the COS-GASS results \citep{bor2015}.  The Stocke-Bowen masses are more than an order of 
magnitude smaller while the COS-Halos masses are 5--40 times larger. 

With most of the gas in the disk, the accretion rate is due to the mass of halo clouds, which have an infall time of $\approx$ 1 Gyr.  That implies an accretion rate from the COS-Halos sample of about 5--90 M$_{\odot}$ yr$^{-1}$, whereas the observed median star formation rate for the COS-Halos sample is 1.3 M$_{\odot}$ yr$^{-1}$.  This would appear to lead to an increase in the mass of the disk with time.  At the lowest accretion rate, the gaseous mass of a typical disk would double in 1--2 Gyr, while at the highest rate, the doubling time is $< 0.1$ Gyr.
However, the gaseous masses of disks have not increased significantly in the past 2.5 Gyr (see \citealt{putman17}), the age difference between the COS-Halos sample and the local universe.  One resolution to this issue is that a larger fraction of the “halo” clouds may be in a rotating disk and not accreting onto the galaxy.  Unfortunately, these galaxies are too distant to be mapped in HI with current synthesis arrays.

\subsection{Where Are the Missing Baryons?}

%On balance, we suggest a working model where about half of all baryons lie within $R_{200}$ and the remainder lies beyond, out to 2--3 $R_{200}$ ($\S$4), with a metallicity of $\sim 0.2$.  Gas beyond the splashback
%radius, which is often 1--2$R_{200}$ \citep{Lau2015,shin16}, has not encountered an accretion shock, so they gas would not be hot unless feedback from supernovae and AGNs has heated the gas and pushed it outward.

The observations are generally consistent with a picture where the extended halo density declines
as $r^{-3/2}$ and where the gas mass within 50 kpc is $\sim 5 \times 10^9$
M$_{\odot}$.  If extrapolated from 50 kpc to larger radii, it implies that the
gas mass increases as $r^{3/2}$, or an increase by a factor of 11 from 50 kpc 
to 250 kpc, about R$_{200}$ for the Milky Way. 
When increased by these amounts, the gas mass of $\sim 6 \times 10^{10}$ is comparable to the
stellar mass, but the gas mass does not account for the missing baryons by 
a factor of 3--10, as discussed above.
The Milky Way type galaxy might be surrounded by warm and neutral
gas with a mass of $\sim 10^{10} M_{\odot}$, and when summed together with the other 
baryonic components is about half of the baryons. 
This raises the question of the location of the missing baryons.

One limiting case is to extrapolate the hot gas and the dark matter distributions until
the ratio becomes the cosmological value.  To do so, we use the NFW
profile to extrapolate the dark matter out to and beyond the virial radius,
which is a slowly increasing function of mass.  For the gas, we assume that
the density law does not change from the inner part, so $M_{gas} \propto r^{3/2}$.  
The reason that we consider this an extreme case is that we might
expect the radial gas distribution to steepen and approach the dark matter 
density relationship at the radius where the cosmological baryon to dark 
matter ratio is reached.

\begin{figure}       %%%%%%%%%%%%% Figure 
\centering
%\epsscale{0.8}
\plotone {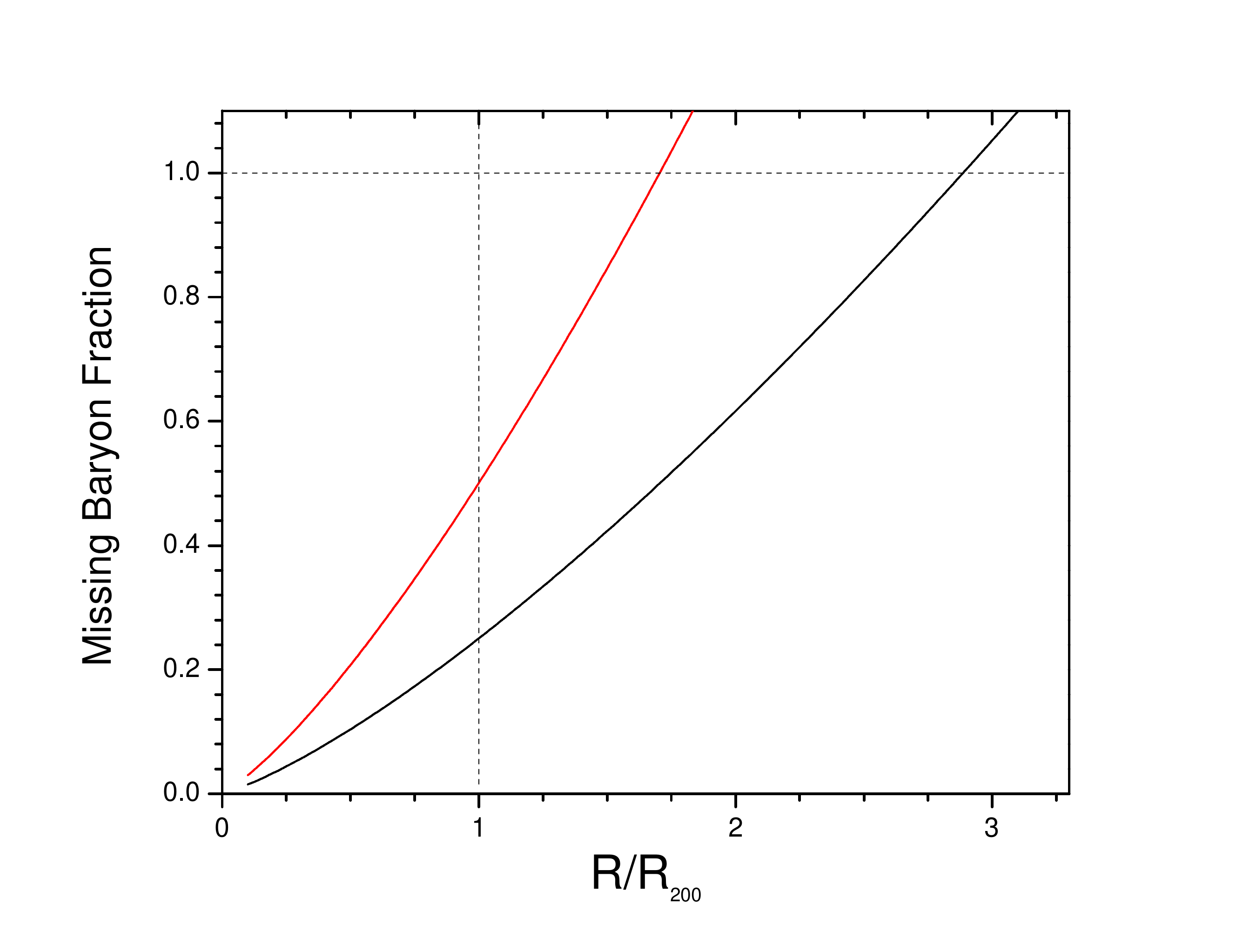}
\caption{
The fraction of the missing baryons relative to the dark matter content 
as a function of radius for the cases where the baryon fraction is 25\% (black curve) and
50\% (red curve) of the cosmic value at the R$_{200}$ (vertical dashed line).  
The horizontal dashed line is the where all missing baryons are accounted for, 
which occurs at 1.7R$_{200}$ and 2.9R$_{200}$ for the two cases. 
}
\label{fig:baryfract}
\end{figure}

Under these assumptions, we calculate the ratio of the increase in the
$M_{gas}$ to the dark matter mass, $M_{DM}$ (Figure ~\ref{fig:baryfract}), 
where we consider the two cases where 25\% and 50\% of the missing baryons are accounted 
for within $R_{200}$, which is fairly typical of our results.  All of the missing
baryons would be accounted by about 1.7--2.9 $R_{200}$.  However, if the radial density
law steepens with radius, the radius containing the missing baryons is moved further outward.
Although 2$R_{200}$ is the classical turnaround radius, dark matter simulations in a 
$\Lambda$CDM universe find the zero velocity surface to typically lie at 4--5 $R_{200}$
for an isolated system \citep{bush2005}.
This is also the typical size of a galaxy group. 
A hot group medium would have a broader radial distribution than for individual
galaxies and that signature may have been detected in O VI and broad Ly$\alpha$ absorption
by \citet{stocke2014}, although the galaxy group interpretation remains inconclusive \citep{stocke17}.  
To determine the fraction of baryons that are bound or unbound to a galaxy will
require a detailed census of the gas mass as a function of temperature.
For gas above about $5 \times 10^5$ K, this will require absorption studies of the
X-ray lines (O VII, O VIII) at a level currently beyond the capabilities of 
current observatories \citep{breg2015}.

\subsection{Future Observations}

There are a number of observational issues to be resolved, such as regarding 
SZ studies and UV absorption line work.  For the SZ work, one would like to consider
the signal as a function of galaxy type or color, provided that there are sufficient
numbers of objects for such an analysis.  Greater angular resolution in SZ studies, such
as from ongoing ground-based work (\textit{ACT}, \citealt{swetz2011}; \textit{SPT}, \citealt{carl2011}), could 
help to determine the size of hot gas region. 
The Stage 4 CMB program, CMB-S4 \citep{CMBs42017} will offer enormous advantages but will 
still require the stacking of galaxies to obtain a detectable signal.
Today, it may be possible to examine nearby galaxies, such as M31 for a SZ signal \citep{taylor2003},
provided that one can make sufficiently accurate Galactic foreground corrections.

From the wealth of UV absorption line data, there are issues relating to absorption from
lower luminosity galaxies near the line of sight, as discussed by \citet{bowen02}. 
This is challenging but necessary work, which does not always lead to a unique result.  

Further observational progress is possible, with the greatest improvements from new instrumentation in the X-ray band.  One goal is to conduct an analog to the UV halo absorption line studies in X-rays for lines that cover a range of temperatures, such as the resonance lines of O VII, O VIII, C V, C VI, and N VII.  This will become possible with an order of magnitude improvement in line detection sensitivity, which will be possible with the proposed Explorer concept \textit{Arcus} \citep{smith2016}, the approved ESA L-class mission \textit{Athena} \citep{barcons2015}, or the NASA large strategic mission concept \textit{Lynx} \citep{gaskin2016}. 
Another goal would be to image the hot coronal gas in X-ray to a significant fraction of $R_{200}$.  This is technically challenging because of the various X-ray backgrounds and long observing times, but \textit{Athena} and \textit{Lynx} should give us important new insights. 

\section{Acknowledgements}
We would like to thank Eric Bell, Oleg Gnedin, Gus Evrard, Jon Miller, Gary Mamon, Arif Babul, 
Chris McKee, Fabrizio Nicastro, John Stocke, Brian Keeney, Yakov Faerman, Bart Wakker, Nicholas Lehner,
Andrey Kravtsov, Joss Bland-Hawthorn
and Richard Mushotzky for their insights and comments.
We are thankful for suggestions and guidance by a patient referee. 
Support for this work is gratefully acknowledged from NASA through ADAP program 
grants NNX16AF23G and NNX15AM93G.

{}

\clearpage

\begin{sidewaystable}[h]
%\begin{table}[h]
\caption[Summary of Hot Halo Properties]{Summary of Hot Halo Properties}
\begin{tabular}{|ccccccc|}
\hline
Galaxy & $L_K (10^{11} L_{\odot})$& Hubble type &  M$_{gas}(<50$ kpc) $(10^9 M_{\odot})$&  M$_{gas}(<R_{vir}$) $(10^{11} M_{\odot})$ & $Z (Z_{\odot})$ & ref \\
\hline \hline
NGC 1961 & $5.2 $  & Sb/c & $5.0_{-0.1}^{+0.2}$   &    $2.3_{-0.9}^{+0.3} { }^{+5.1}$ & 0.5 & AB11\\
NGC 1961  &  $5.2 $  & Sb/c & $11$ & $11$ & 0.12 (measured) B13\\
NGC 6753  &   $3.9$  & Sb&  12 & 5.0  & 0.13 (measured) &B13\\
UGC 12591 & $5.6 $ & S0/a & $4.4^{+0.7}_{-0.9}$ & $1.5_{-1.3}^{+1.6} { }^{+3.0}$ & 0.5 &DAB12 \\
stacked luminous & $1.4^{+0.8}_{-0.4} $ &-& $4.1^{+0.6}_{-1.0} (\text{stat}) \pm2.9 \text{(sys)}$ & - & 0.3 &AB13a\\
stacked faint& $0.3^{+0.4}_{-0.1}$ & - &$ 0.9^{+0.5}_{-0.4} (\text{stat}) \pm0.6 \text{(sys)}$* &-& 0.3 &AB13a\\
stacked early-type& $0.7^{+1.2}_{-0.3}$ &E and S0& $1.9\pm0.9 (\text{stat}) \pm1.3 \text{(sys)}$ &-& 0.3 AB13a\\
stacked late-type& $0.4^{+1.0}_{-0.2}$ &Sabc \& Irr& $1.2^{+0.5}_{-0.6} (\text{stat}) \pm0.8 \text{(sys)}$* &-& 0.3 AB13a\\
NGC 720 & $1.6 $& E5 & $8\pm1.5$ & $1.6\pm 0.5$ &$\approx0.6$ (measured) & H06\\
NGC 720 &$1.6$ & E5 & $11\pm2$ & $2.7\pm0.6$ &$\approx0.6$ (measured) & H11\\
NGC 720 & $1.6$ & E5 & $6.5\pm0.5$ & $0.8\pm0.1$ &0.6 AB13b\\
NGC 1521 & $2.3$ & E3& 10 & 6 &$\approx 0.4$&H12\\ 
\hline
\end{tabular}\\
\scriptsize{Recent measurements of hot halo masses. For each column, $L_K$ comes from 2MASS and the Hubble type from NED. Best-fit measurements of the hot halo gas mass within 50 kpc and within the virial radius are then listed for each galaxy. In general, the mass within 50 kpc is fairly secure, while the mass within the virial radius depends on extrapolating the density profile out to much larger radii than it is observed. The errors quoted are statistical errors based on uncertainties in the surface brightness profile; other sources of error are discussed in the references for each paper (most of these other errors are folded into the systematic errors quoted for the stacked galaxies, however). Note that the asterisks on the stacked faint galaxies and the stacked late-type galaxies denote lower confidence that we are actually detecting and characterizing extended emission around these galaxies. For our measurements NGC 1961 and UGC 12591, the second listed uncertainty on the mass within the virial radius accounts for the possibility of a flattened profile, as discussed in the text. For the stacked galaxies, we do not extrapolate the mass to the virial radius since we do not have a strong measurement of the slope of the density profile. The second-to-last column lists the assumed (or measured, in some cases) metallicity used to convert the surface brightness profile into a gas mass (for NGC 720 and NGC 1251 a metallicity profile is measured, and we quote an approximate luminosity-weighted average). The references in the final column are: AB11 = \citet{ander2011};  B13 = \citet{bogdan13}; DAB12 = \citet{Dai2012}; AB13a = \citet{ander2013}; AB13b = \citet{anderson2014}; H06 = \citet{humph06}; H11 = \citep{humph11}; H12 = \citet{humph12} }
%\end{table}
\end{sidewaystable}

\begin{table}[h]
%\begin{sidewaystable}%[h]
\caption[Halo Cool-Warm Gas Masses]{Halo Cool-Warm Gas Masses}
\begin{tabular}{|c|c|c|c|c|c|}
\hline 
Galaxy/Survey & R$_{max}$ & M(HI, R$_{max}$) & M$_{gas}$(R$_{max}$) & M$_{gas}$(R$_{200}$) & Comments \\ 
\hline 
• & (kpc) & (M$_{\odot}$) & (M$_{\odot}$) & (M$_{\odot}$) & • \\ 
\hline 
Milky Way  & 15 & 6E7 & 2.8E8 & 1.5E9 & T $<1 \times 10^6$ K, \citet{zheng15} \\ 
\hline 
Andromeda & 50 & 4E7 & 1.2E9 & 2.8E9 & \citet{lehner15} \\ 
\hline 
Stocke-Bowen; halo & 150 & 1E5 & 5E7 & • & this work \\ 
\hline 
S-B disk & 53 & 7E6 & 5E7 & • & this work \\ 
\hline 
S-B total & 150 & 7.1E6 & 1E8 & 1.1E8 & this work \\ 
\hline 
COS-Halos; halo & 150 & 1E7 & 5E9 &   & this work; 0.2 solar \\ 
\hline 
COS-H; disk & 50-100 & 1.2-5E9 & 7-30E9 & • &  this work \\ 
\hline 
Cos-H; total & 150 & 1.2-5E9 & 1.1E10 & 1.2E10 &  this work \\ 
\hline 
Cos-H 2017 & 150 &   & 8.4E10 & 8.8E10 & \citet{proc2017} \\ 
\hline 
\end{tabular} \\
\end{table}
%\end{sidewaystable}%[h]

\end{document}